\documentclass[aps,prb,twocolumn,groupaddress,floatfix]{revtex4}

\usepackage{graphicx}
\usepackage{dcolumn}
\usepackage{bm}
\usepackage{amsmath}
\usepackage{multirow}
\usepackage{xcolor}
\usepackage{simpler-wick}
\usepackage{gensymb}
\usepackage{enumitem}

\newcommand{\tbhc}[1]{\multicolumn{1}{c}{\textbf{#1}}}

\newcommand{\tbhr}[1]{\multicolumn{1}{r}{\textbf{#1}}}

\newcolumntype{f}[1]{D{.}{.}{#1}}

\begin{document}

\title{Real-time coupled-cluster approach for the cumulant Green's function}
\date{\today}

\author{F. D. Vila}
\author{J. J. Rehr}
\author{J. J. Kas}
\affiliation{Department of Physics, University of Washington, Seattle, WA 98195}
\author{K. Kowalski}
\affiliation{William R. Wiley Environmental Molecular Sciences Laboratory, Battelle, Pacific Northwest National Laboratory, K8-91, P. O. Box 999, Richland, Washington 99352}
\author{B. Peng}
\affiliation{Physical Sciences Division, Pacific Northwest National Laboratory, Richland, WA 99354}

\begin{abstract} 

Green's function methods within many-body perturbation theory provide a
general framework for treating electronic correlations in excited states. 
Here we investigate the cumulant form of the one-electron Green's
function based on the coupled-cluster equation of motion approach in an
extension of our previous study. The approach yields a non-perturbative
expression for the cumulant in terms of the solution to a set of
coupled first order, non-linear differential equations. The method thereby
adds non-linear corrections to traditional cumulant methods linear in
the self energy.  The approach is applied to the core-hole Green's function 
and illustrated for a number of small molecular systems.
For these systems we find that the non-linear contributions lead to
significant improvements both for quasiparticle properties such as
core-level binding energies, as well as the satellites 
corresponding to inelastic losses observed in photoemission spectra.

\end{abstract}

\date{\today}

\maketitle

\section{Introduction}
\label{sec:intro}

Wavefunction based coupled-cluster (CC) methods have traditionally been
used for accurate calculations of ground state electronic correlation effects
in molecular systems.\cite{Bartlett2009} In contrast, Green's function (GF)
methods within many-body perturbation theory (MBPT) provide a general
formalism for treating electronic correlation effects in
excited states.\cite{Hedin99review}
These effects include the quasiparticle line shape as well as satellites
corresponding to the intrinsic inelastic losses due to many-body excitations
that are observed in x-ray photoemission spectra (XPS).
Recently, a number of CC approaches have been developed for the one-particle
GF. For example, a perturbation theoretic approach for
the CCGF in frequency-space has been developed by
Peng and Kowalski.\cite{PengKowalski2016,PengKowalski2018}
As an alternative, our treatment here is based on the equation-of-motion
coupled-cluster (EOM-CC) approach for the
cumulant GF in real-time.
Cumulant GF formulations have proved to be advantageous for
understanding correlation properties in
condensed matter,\cite{Hedin99review,sky}
and are exact for model systems with electrons coupled to bosons due to
the linked-cluster theorem.\cite{langreth70}
Our initial development was restricted to the treatment of the
core-hole GF with an approximate Hamiltonian for a system
with a deep core-hole.\cite{RVKNKP}
Here the approach is extended to include all terms within the
CC-singles approximation. This extension yields an exact expression for the
cumulant in terms of the solution to a set of coupled first order
non-linear differential equations. 
In particular this approach builds in non-linear terms in the cumulant
which significantly improve quantitative calculations.
As a quantitative test, the method is applied to a number
of small molecular systems.  We find that the approach including
non-linear corrections yields quasiparticle properties to high accuracy,
as well as an approximate treatment of satellites. 
Moreover, the approach converges rapidly
and gives very good results even at the leading order
truncations of the EOM-CC that include non-linear terms.

The structure of this paper is as follows: In Sec \ref{sec:theo}.\
we first derive an expression for the cumulant in a spin-orbital basis in
terms of the one-particle self-energy,
following the approach of Aryasetiawan et al.\cite{PhysRevLett.77.2268}
We then show that the 2nd-order approximation
for the self-energy can be derived within the CC approximation
using perturbation theory.
Next we develop a EOM-CC approach for the cumulant that includes all terms
in the CC-singles approximation. We show that this yields
a direct, non-perturbative relation between the CC GF 
and the exponential cumulant representation in real-time in terms
of a set of non-linear differential EOM.
Results are then presented in Sec. \ref{sec:res}. for the quasiparticle properties and 
satellites in the spectral function for a number of
small molecular systems.
Finally we present a brief summary and conclusions.

\section{Theory}
\label{sec:theo}

\subsection{Retarded Green's function and cumulant in a spin-orbital basis set}
\label{sec:gfcum}

Within a basis of single-particle spin-orbitals \{$\phi_p(r)\}$
the retarded Green's functions in the time domain is
defined as (see SI Sec.\ I.\ A.)
\begin{equation}
\label{eqn:gpq}
G_{pq}(t,t') =
-i \Theta(t-t')
\left<0\left| \left\{a_p(t), a_q^\dagger(t') \right\} \right| 0 \right>,
\end{equation}
where the creation and annhilation operators $a_p^\dagger(t)$ and $a_q(t')$ are
associated with spin-orbitals $p$ and $q$ respectively.
Setting $t'=0$ for convenience with time-dependent operators 
$O(t) = e^{iHt}Oe^{-iHt}$, where $H$ is the
full $N$-particle Hamiltonian of the system, we obtain
\begin{equation}
\label{eqn:gpqt}
\begin{split}
G_{pq}(t) =
-i \Theta(t)[ &
  e^{iE_0t} \left<0\left| a_p e^{-iHt} a_q^\dagger \right| 0 \right>  \\
+ & e^{-iE_0t}\left<0\left| a_q^\dagger e^{iHt} a_p \right| 0 \right> ],
\end{split}
\end{equation}
where $E_0$ is the total energy of the ground state $|0\rangle$,
which is the eigenvalue of $H \left| 0 \right> = E_0 \left| 0 \right>$.
From Eq.\ (\ref{eqn:gpqt}) we obtain the Green's function
of the non-interacting system $G^0_{pq}(t)$ using the relations
\begin{eqnarray}
H   &\rightarrow& H_0 = \sum_p\epsilon_p a_p^{\dagger}a_p \nonumber \\
\left| 0 \right> &\rightarrow& \left| \Phi \right> \nonumber \\
E_0 &\rightarrow& E^0_0 =
\left< \Phi \left| H_0 \right| \Phi \right> = \sum_i \epsilon_i.
\end{eqnarray}
Here and below we use the convention where $i, j, k, ...$
correspond to occupied (``hole'') states in the Hartree-Fock single-determinant $\left| \Phi \right>$, $a, b, c, ...$ indices correspond to unoccupied (``particle'') states, and $p, q, r, ...$ can be either.
With these definitions,
\begin{equation}
\label{eqn:g0diag}
G^0_{pq}(t) = -i\Theta(t) e^{-i\epsilon_p t}\delta_{pq}.
\end{equation}
The cumulant ansatz for the Green's function in the time-domain for a given
one-particle orbital $p$ is
defined\cite{PhysRevB.90.085112,PhysRevLett.77.2268} by the exponential representation
\begin{equation}
\label{cumold}
G_{p}(t) = G^0_{p}(t)e^{C_{p}(t)}.
\end{equation}
Here $G_{p}(t)$ and $G^0_{p}(t)$ are the interacting and non-interacting retarded
Green's functions for one-particle orbital $p$, respectively,
and $C_{p}(t)$ is
the retarded cumulant. Eq.\ (\ref{cumold})
can be generalized to its non-diagonal, spin-orbital matrix
form similar to Eq.\ (\ref{eqn:gpq}) (see SI Sec.\ I.\ A.):
$\hat{G}(t) = \hat{G}^0(t)e^{\hat{C}(t)}$,
where $\hat{G}$, $\hat{G^0}$ and $\hat{C}$ are the matrix forms of the Green's
functions and cumulant respectively.
In the conventional formulation based on the decoupling approximation,
the cumulant is obtained
by matching the cumulant expansion of $G_p(t)$ to first order
with that from the Dyson equation $G=G^0+G^0\Sigma G$,
(see SI Sec.\ I.\ B.).\cite{PhysRevB.90.085112,PhysRevLett.77.2268}
Since $\hat{G}^0$ is diagonal in the one-particle HF
eigenstates $p$, the cumulant satisfies the relation
\begin{equation}
G^0_{pp}(t) C_{pq}(t) = \int dt_1 dt_2 G^0_{pp}(t-t_1)
\Sigma_{pq}(t_1-t_2) G^0_{qq}(t_2).
\end{equation}
Introducing the Fourier transform of the double convolution
in the right hand term and the form
of $\hat{G}^0$ from Eq.\ (\ref{eqn:g0diag}) we obtain for $t\geq0$
\begin{equation}
C_{pq}(t) = i \int \frac{d\omega}{2\pi} e^{i(\omega+\epsilon_p) t}
G^0_{pp}(\omega) \Sigma_{pq}(\omega) G^0_{qq}(\omega),
\end{equation}
Then shifting the $\omega$ integration variable in the Fourier transform 
with $G^0_{pp}(\omega)=(\omega-\epsilon_p+i\delta)^{-1}$,
and approximating the self-energy with its diagonal form
$\Sigma_{pq}(\omega) \simeq \Sigma_{pp}(\omega) \delta_{pq}$,
the diagonal elements of the cumulant become
\begin{equation}
\label{eqn:cum_diag_t}
C_{pp}(t) = \int \frac{d\omega}{2\pi} \frac{i\Sigma_{pp}(\omega+\epsilon_p)}
{(\omega+i\delta)^2} e^{-i\omega t}.
\end{equation}
Finally since the matrix exponential is now diagonal 
\begin{equation}
\label{eqn:g_cum_diag}
G_{pp}(t) = -i \Theta(t) e^{-i \epsilon_p t + C_{pp}(t)},
\end{equation}
which is the standard form of the cumulant Green's function for
the diagonal elements.

\subsection{Second order approximations}
\label{sec:2ose_cum}

As shown above, the cumulant to lowest order 
is linear in the one-particle self-energy, so the problem reduces to finding
a suitable approximation for $\Sigma_{pp}(\omega)$.
In the GW approximation of Hedin,\cite{Hedin99review} for example,
$\Sigma = iGW$ is approximated to first order in the screened
electron-electron interaction $W=\epsilon^{-1}V$, where $\epsilon$ is the 
dielectric function. Here, following the usual practice
for molecular systems, screening is neglected, and hence to leading order
$\Sigma=\Sigma^{(2)}$ becomes the 2nd-order
self-energy (SE2).\cite{linderberg2004propagators,szabo1996modern}
We also show that $\Sigma^{(2)}$
can obtained using the perturbation theory and the
CC approximation with, at least doubles, to lowest non-vanishing order.

\subsubsection{Second order self-energy and cumulant}

Within MBPT and assuming a single-determinant Hartree-Fock reference
$\left|\Phi\right>$, the 2nd-order 
self-energy in the spin-orbital basis is given by\cite{linderberg2004propagators,szabo1996modern}
\begin{equation}
\label{eqn:2ose}
\begin{split}
\Sigma_{pq}^{(2)}(\omega) =&
\frac{1}{2} \sum_{iab}
\frac{\left< pi \left| \right| ab \right> \left< ab \left| \right| qi \right>}
{\omega+\epsilon_{i}-\epsilon_{a}-\epsilon_{b}} + \\
&+\frac{1}{2} \sum_{ija}
\frac{\left< pa \left| \right| ij \right> \left< ij \left| \right| qa \right>}
{\omega+\epsilon_{a}-\epsilon_{i}-\epsilon_{j}},
\end{split}
\end{equation}
where $\left< pq \left| \right| rs \right> = \left< pq | rs \right> - \left< pq
| sr \right>$ are the antisymmetric Coulomb integrals over the real $p, q,
r, s$ spin-orbitals.
The diagonal terms in Eq.\ (\ref{eqn:cum_diag_t}) can be written as
\begin{equation}
\label{eqn:2oseatp}
\Sigma_{pp}^{(2)}(\omega+\epsilon_p) =
\frac{1}{2} \sum_{iab}
\frac{\left< pi \left| \right| ab \right>^2}
{\omega-\epsilon_{pi}^{ab}} + \\
\frac{1}{2} \sum_{ija}
\frac{\left< pa \left| \right| ij \right>^2}
{\omega-\epsilon_{pa}^{ij}},
\end{equation}
where
$\epsilon_{pi}^{ab} = \epsilon_{a}+\epsilon_{b}-\epsilon_{p}-\epsilon_{i}$
and
$\epsilon_{pa}^{ij} = \epsilon_{i}+\epsilon_{j}-\epsilon_{p}-\epsilon_{a}$.
Consequently to 2nd-order in perturbation theory the 2nd order
cumulant is obtained with
$\Sigma_{pp} \simeq \Sigma_{pp}^{(2)}$
in Eq.\ (\ref{eqn:cum_diag_t})
\begin{equation}
\label{eqn:cum_t_2ose}
\begin{split}
C^{(2)}_{pp}(t) =&
\frac{1}{2} \sum_{iab} \left< pi \left| \right| ab \right>^2
\int \frac{d\omega}{2\pi}
\frac{i e^{-i \omega t}}
{\omega^2\left(\omega-\epsilon_{pi}^{ab}\right)} + \\
+& \frac{1}{2} \sum_{ija} \left< pa \left| \right| ij \right>^2
\int \frac{d\omega}{2\pi}
\frac{i e^{-i \omega t}}
{\omega^2\left(\omega-\epsilon_{pa}^{ij}\right)}.
\end{split}
\end{equation}
Then using the identity
\begin{equation}
\int \frac{d\omega}{2\pi}
\frac{i e^{-i \omega t}}
{\omega^2\left(\omega-\epsilon\right)} = \frac{1}{\epsilon^2}
\left(e^{-i \epsilon t} +i \epsilon t -1 \right) \mathrm{sgn}(t)
\end{equation}
the 2nd-order cumulant can be expressed as
\begin{equation}
\label{eqn:cum_coupl}
\begin{split}
C^{(2)}_{pp}(t) =&
\frac{1}{2}\sum_{iab} \left(u_{ip}^{ab}\right)^2
\left(e^{-i \epsilon_{pi}^{ab} t} +i \epsilon_{pi}^{ab} t -1 \right)  + \\&
+\frac{1}{2}\sum_{ija} \left(u_{pa}^{ij}\right)^2
\left(e^{-i \epsilon_{pa}^{ij} t} +i \epsilon_{pa}^{ij} t -1 \right)
\end{split}
\end{equation}
where the cumulant amplitudes $u_{pq}^{rs}$ are 
\begin{equation}
\label{}
u_{pq}^{rs} =
\frac{\left< pq \left| \right| rs \right>}
{\epsilon_{pq}^{rs}}.
\end{equation}
For an occupied core state $p$, the $u_{ip}^{ab}$ coefficients are equivalent to the doubly-excited CC amplitudes approximated to 1st order in Moller-Plesset
MBPT.\cite{crawford2000introduction}
 This result is equivalent to the first iteration in the solution of any CC formulation that includes T$_2$ (when the initial guess is the null vector),\cite{crawford2000introduction} thus demonstrating a direct connection between the exponential form of the retarded cumulant and the CC approach. The behavior of the cumulant
for the 2nd-order self energy is similar to that for electrons coupled to
bosonic excitations labeled by an index $q$ in the quasi-boson approximation
with coupling coefficients $g_q$,\cite{Hedin99review} 
\begin{equation}
C^{(2)}_{pp}(t) = \sum_{q} \frac {g_q^2}{\omega_q^2}
\left(e^{-i \omega_q t} +i \omega_q t -1 \right)
\end{equation}
where $g_q =\omega_q u_q$.

For analysis purposes, it is convenient to define a cumulant
kernel $\beta(\omega)$ that characterizes the spectrum of excitations
\begin{equation}
\beta(\omega) = \sum_q g_q^2 \delta(\omega-\omega_q) ,
\end{equation}
where for the SE2,
\begin{equation}
\beta(\omega) = -\frac{1}{\pi}\, {\rm Im}\, \Sigma^{(2)}_{pp}(\omega+\epsilon_p).
\end{equation}
Thus $\beta(\omega)$ is given by the poles of the $\Sigma^{(2)}_{pp}$.
As a consequence the cumulant can also be defined by the kernel $\beta(\omega)$
\begin{equation}
C^{(2)}_{pp}(t) = \int d\omega\, \frac{\beta(\omega)}{\omega^2}
\left(e^{-i \omega t} +i \omega t -1 \right).
\end{equation}
This expression is referred to as the Landau form of the cumulant, and
facilitates the interpretation of excitations in the spectrum.
For example, since $C^{(2)}_{pp}(0)=C^{(2)'}_{pp}(0)=0$, the Landau form
guarantees that the spectral function
$A_{pp}(\omega)=(-1/\pi)\,{\rm Im}\, G_{pp}(\omega)$
is normalized and has an invariant centroid at the independent particle energy
$\epsilon_p$ as in Koopmans' theorem.

\subsubsection{2nd-order CC Green's function}
\label{sec:ccg_t_pt}

It is interesting to note that 2nd-order perturbation theory
for $G^R_{pq}(\omega)$ based on the CC ansatz and
including at least T$_2$ double excitations yields the same 2nd-order
self-energy $\Sigma^{(2)}(\omega)$ discussed above.
Here we demonstrate this equivalence in the
frequency domain, since this treatment avoids
the complication of expanding the exponential propagation operator
in Eq.\ (\ref{eqn:gpqt}) in successive orders of perturbation theory.
We start with the frequency domain version of the GF corresponding to Eq.\ (\ref{eqn:gpqt}):
\begin{equation}
\label{eq:gpqrw1}
\begin{split}
G^R_{pq} (\omega) =&
\left\langle \Phi \left| (1+\Lambda) \bar{a^{\dagger}_q}
(\omega + \bar{H}_N + i\delta)^{-1} \bar{a_p} \right| \Phi \right\rangle +\\
&\left\langle \Phi \left| (1+\Lambda) \bar{a_p}
(\omega - \bar{H}_N + i\delta)^{-1} \bar{a^{\dagger}_q} \right| \Phi \right\rangle,
\end{split}
\end{equation}
equivalent to the CC GF (see SI Sec.\ I C) in Eq.\ (17) of
Ref.\ \onlinecite{PhysRevA.94.062512}, but including both the $N-1$ and $N+1$ branches. We include the $N+1$ branch for completeness, although in the cases studied here it only has a small contribution to the total GF.
In Eq. \ref{eq:gpqrw1}, $|\Phi \rangle$ is the reference HF determinant, $\bar{O} = e^{-T}Oe^{T}$ is the similarity transformed $O$
operator, $H_N$ is the normal ordered Hamiltonian,
$\Lambda$ is the CC de-excitation operator, and we have used the
Baker-Campbell-Hausdorff (BCH) relation 
\begin{equation}
\label{eq:apbar}
\begin{split}
\bar{a}_p =&  a_p + [a_p,T] \\
\bar{a^{\dagger}_q} =& a^{\dagger}_q + [a^{\dagger}_q,T].
\end{split}
\end{equation}
Inserting the retarded form of the auxiliary $N-1$ operator $X_{p}(\omega) = (\omega+\bar{H}_N+i\delta)^{-1} \bar{a}_p$ 
and the $N+1$ operator $Y_{q}(\omega) = (\omega-\bar{H}_N+i\delta)^{-1} \bar{a^{\dagger}_q}$
into Eq.\ (\ref{eq:gpqrw1})
\begin{equation}
\label{eq:gpqrw2}
\begin{split}
G^R_{pq} (\omega) =&
\left< \Phi \left| (1+\Lambda) \bar{a^{\dagger}_q}
X_{p}(\omega) \right| \Phi \right> +\\
&\left< \Phi \left| (1+\Lambda) \bar{a}_p
Y_{q}(\omega) \right| \Phi \right>.
\end{split}
\end{equation}
If we now assume that $T=T_2=\frac{1}{4}\sum_{ijab}t_{ij}^{ab} a^{\dagger}_a a^{\dagger}_b a_j a_i$ and $\Lambda=\Lambda_2=\frac{1}{4}\sum_{ijab}\lambda_{ij}^{ab} a^{\dagger}_i a^{\dagger}_j a_b a_a$,
and expand all the operators to 2nd-order and simplify (see SI Sec.\ I D)
\begin{equation}
\label{eq:g2pqrw2}
\begin{split}
G^{R(2)}_{pq} (\omega) =& \frac{\delta_{pq}}{(\omega-\epsilon_p)} +
\frac{1}{(\omega-\epsilon_p)} \times \\
&\times\left[ \frac{1}{2}\sum_{ija} \frac{v^{qa}_{ij}v^{pa}_{ij}}
{(\omega+\epsilon_a-\epsilon_i-\epsilon_j)}+ \right.\\
&+\left.
\frac{1}{2}\sum_{iab} \frac{v^{pi}_{ab}v^{qi}_{ab}}
{(\omega+\epsilon_i-\epsilon_a-\epsilon_b)}
\right]\frac{1}{(\omega-\epsilon_q)}.
\end{split}
\end{equation}
Note that this 2nd-order result for Green's function has the
form of a Dyson equation
$\hat{G}^{R(2)}(\omega) = \hat{G}^{R}_0(\omega) + \hat{G}^{R}_0(\omega)
\hat{\Sigma}^{(2)}(\omega)\hat{G}^{R(2)}_0(\omega)$, where the expression
in brackets is identical to the 2nd-order self-energy
of Eq.\ (\ref{eqn:2ose}).
Consequently  the Fourier transform of Eq.\ (\ref{eq:g2pqrw2}) yields
the same 2nd-order cumulant as Eq.\ (\ref{eqn:cum_diag_t}).

\subsection{Real-time EOM-CC Cumulant GF}
\label{sec:rt_core_gf}

In order to explore corrections to the 2nd-order approximation
for the cumulant GF, we  now develop a real-time approximation
based on the more general equation of motion (EOM-CC) ansatz.
Our treatment here extends that introduced in our original
approach\cite{RVKNKP} by including terms up to third order in the
CC amplitudes, and gives a non-perturbative representation for the cumulant.
For definiteness, we restrict our discussion here to the retarded
core-hole Green's function for a given deep core level $p=c$, $G^R_c=G_{cc}$
given by
\begin{equation}
\label{eqn:grpqt_a}
\begin{split}
G_{c}^{R}(t) =
-i \Theta(t) &
  e^{iE_0t} \left<0\left| a_c e^{-iHt} a_c^\dagger \right| 0 \right> + \\
-i \Theta(t) &
e^{-iE_0t}\left<0\left| a_c^\dagger e^{iHt} a_c \right| 0 \right>.
\end{split}
\end{equation}
We then introduce the separable approximation to the ground state $\left| 0
\right> \simeq a_c^\dagger \left| N-1 \right>$, where $\left| N-1 \right>$ is
the fully correlated $N-1$ electron part of the $N$ electron wavefunction with
the core electron separated from it. Inserting this into Eq.\
(\ref{eqn:grpqt_a}) and remembering that $a_c a_c = 0$,
\begin{equation}
\label{eqn:grpqt_b}
G_{c}^{R}(t) =
-i \Theta(t) e^{-iE_0t}\left<N-1\left| e^{iHt} \right| N-1 \right>.
\end{equation}
Formally $\left| N-1, t \right> = e^{iHt}\left| N-1 \right>$ is a
solution to
\begin{equation}
-i \frac{d\left| N-1, t \right>}{dt} = H \left| N-1, t \right>,
\end{equation}
so that 
\begin{equation}
\label{eqn:grpqt_c}
G_{c}^{R}(t) =
-i \Theta(t) e^{-iE_0t}\left<N-1\right| \left. N-1, t \right>.
\end{equation}
Next we assume a time-dependent, CC ansatz for
$\left| N-1, t \right> = N(t) e^{T(t)} \left| \phi \right>$. It is important to
note that the excitation operator $T$ acts in the $N-1$ particle Fock space, rather than in the $N$ space as in standard ground-state CC, and that the reference determinant is
$\left| \phi \right> = a_c \left| \Phi \right>$,
where $\left| \Phi \right>$ is the $N$ electron HF
determinant of the ground state.
Thus, our treatment here refers to a CC approximation to
the excited states involved in the calculation of the Green's function,
rather that the
typical applications where the CC ansatz is used for the ground state.
Inserting this ansatz into
the differential equation for $\left| N-1, t \right>$ and left multiplying 
by $e^{-T(t)}$, we obtain the coupled EOM
\begin{equation}
\label{eqn:modsch}
-i \left[ \frac{d \ln N(t)}{dt} + \frac{d T(t)}{dt} \right] =
\bar{H}(t) \left| \phi \right>,
\end{equation}
where the similarity transformed Hamiltonian is
$\bar{H}(t) = e^{-T(t)}He^{-T(t)}$. Here in order to simplify both the notation
and the computation of the matrix elements, instead of the
exact second-quantized Hamiltonian
\begin{equation}
H = \sum_{pq} h_{pq} a_p^\dagger a_q + 
\frac{1}{4} \sum_{pqrs}
v_{pq}^{rs} a_p^\dagger a_q^\dagger a_s a_r,
\end{equation}
where $h_{pq}$ are the single particle kinetic and electron-nuclei molecular
orbital integrals, we introduce its normal ordered form $H_N
= H - \left< \phi \left| H \right| \phi \right>$. The similarity transformed
form of $H_N$ is $\bar{H}_N(t) = \bar{H}(t) -
E^{N-1}$, where $E^{N-1} = \left< \phi \left| \bar{H} \right|
\phi \right>$. It should be noted that with an $N-1$ reference, 
$H_N$ no longer has the usual diagonal single particle term
but rather
\begin{equation}
H_N = \sum_{pq} f_{pq} \left\{ a_p^\dagger a_q\right\}' + 
\frac{1}{4} \sum_{pqrs}
v_{pq}^{rs} \left\{ a_p^\dagger a_q^\dagger a_s a_r\right\}',
\end{equation}
where the $\{\}'$ is a reminder that the normal ordering is done with respect
to the $N-1$ particle reference $\left| \phi \right>$, and $f_{pq} = \epsilon_p \delta_{pq} - v_{pc}^{qc}$.
We now follow the usual CC approach of
projecting Eq.\ (\ref{eqn:modsch}) from the left with reference $\left< \phi \right|$ and the $i \rightarrow a,j \rightarrow b,...$ excited reference
$\left< \phi_{ij...}^{ab...} \right|$ to separate the EOM for
$N(t)$ and $T(t)$,
\begin{equation}
\label{eq-dlnndt}
-i \frac{d \ln N(t)}{dt} = \left< \phi \left| \bar{H}_N(t) \right| \phi \right>
+ E^{N-1},
\end{equation}
\begin{equation}
\label{eq-dtdt}
-i \left< \phi_{ij...}^{ab...} \left| \frac{d T(t)}{dt} \right| \phi \right> =
\langle \phi_{ij...}^{ab...} \left| \bar{H}_N(t) \right| \phi \rangle.
\end{equation}
It is interesting to note that these equations have a structure
identical to the standard CC equations for the ground state,\cite{crawford2000introduction}
 except that now we are interested in $N(t)$ and $T(t)$ instead of $E_{CC}$
and $T$. Moreover, the equations are now non-linear coupled  first order 
differential equations rather than algebraic equations. The only
matrix elements required to get explicit
expressions are $\left< \phi \left| \bar{H}_N(t) \right| \phi \right>$,
$\left< \phi_{ij...}^{ab...} \left| {d T(t)}/{dt} \right| \phi \right>$,
and $\left< \phi_{ij...}^{ab...} \right| \bar{H}_N(t) \left| \phi \right>$.
In order to evaluate these results,
we need to introduce some further approximations.
First, we assume that the ground state is uncorrelated, i.e.
$\left| N-1 \right> \simeq a_c \left| \Phi \right> = \left| \phi \right>$,
so that
\begin{equation}
\begin{split}
\left<N-1\right| \left. N-1, t \right> =&
N(t) \left<N-1\left| e^{iHt}\right| \phi \right> \\
=&N(t) \left< \phi \left| e^{iHt}\right| \phi \right> \\
=&N(t) \left( 1+ \left< \phi \left| R(t) \right| \phi \right> \right) \\
=&N(t).
\end{split}
\end{equation}
Here $R(t)$ is the excitation operator that collects all excited terms arising from the series expansion of $e^{iHt}$ and has expectation value $\left< \phi \left| R(t) \right| \phi \right> = 0$ due to orthogonality.
The approximation of an uncorrelated ground state
could be relaxed by replacing $|\Phi\rangle$ by $\exp(T)|\Phi\rangle$.
However, with the approximation introduced above, 
the core-hole Green's function $G_c^Rt)$ 
is directly related to the normalization factor $N(t)$
\begin{equation}
G_{c}^{R}(t) =
-i \Theta(t) e^{-iE_0t}N(t).
\end{equation}
We note that the normalization factor $N(t)$ also corresponds to
the vacuum fluctuations in field theory treatments.\cite{NozieresDeDominicis}
Given that no correlation is included
in the ground state, we can simply approximate $E_0 
\simeq E_\mathrm{HF}$, i.e., the HF energy of the $N$-particle system,
and hence $E_0-E^{N-1} \simeq \epsilon_c$, as expected
from Koopmans' theorem.

The logarithmic derivative in the EOM in Eq.\ (\ref{eq-dlnndt}) implies that $N(t)$ is a pure exponential, so $G_c^R(t)$ has an explicit cumulant form in the time-domain
\begin{equation}
\label{eqn:cum_form}
G_{c}^{R}(t) = -i \Theta(t) e^{-i\epsilon_c t} e^{C_c^{R}(t)},
\end{equation}
where the cumulant is obtained by integrating Eq.\ (\ref{eq-dlnndt})
\begin{equation}
\label{eqn:cum_t_f}
C_c^{R}(t) = i \int_0^t \left< \phi \left| \bar{H}_N(t) \right| \phi\right> dt'.
\end{equation}
with the boundary condition $C_c^R(0)=0$.
We then make one further approximation for the treatment here,
namely that the operator $T$ is restricted to
single excitations $T(t) = T_1(t) \equiv
\sum_{ia} t_i^a(t) \{a_a^\dagger a_i \}'$, where again the $\{\}'$ make
explicit that the contractions are with respect to that $N-1$ reference.
Thus the occupied indices $i, j, ...$ do not include the the core index $c$, though the unoccupied $a, b, ...$ levels do. This will be assumed implicitly in all sums below. We also suppress the time-dependence label in the CC amplitudes $t_i^a(t)$ unless needed for clarity. In order to obtain explicit expressions for the amplitudes within this $T_1$ approximation we need to calculate the matrix elements $\left< \phi \left| \bar{H}_N(t) \right| \phi \right>$,
$\left< \phi_{i}^{a} \left| {d T(t)}/{dt} \right| \phi \right>$,
and $\left< \phi_{i}^{a} \right| \bar{H}_N(t) \left| \phi \right>$.
To begin, we note that although the full similarity transformed Hamiltonian
is given by
\begin{equation}
\begin{split}
\bar{H}_N(t) =& H_N + \left(H_N T_1(t)\right)_c +
\frac{1}{2!} \left(H_N T_1(t)^2\right)_c + \\
&\frac{1}{3!} \left(H_N T_1(t)^3\right)_c +
\frac{1}{4!} \left(H_N T_1(t)^4\right)_c,
\end{split}
\end{equation}
the quartic terms do not contribute to the matrix elements of
interest. After some straightforward, though tedious algebra and diagrammatic
analysis, we obtain a compact expression for Eq.\ (\ref{eq-dlnndt})
\begin{equation}
\label{eqn:matel1}
\begin{split}
-i\frac{d C_c^R(t)}{dt} &= \left< \phi \left| \bar{H}_N(t) \right| \phi \right> \\
&= \sum_{ia} f_{ia} t_i^a +
\frac{1}{2} \sum_{ijab} v_{ij}^{ab} t_j^b t_i^a,
\end{split}
\end{equation}
The EOM of the CC amplitudes are
\begin{equation}
\label{eq:rt_eom_ccs}
-i \dot {t}_i^a = \left< \phi_{i}^{a} \left| \frac{d T(t)}{dt} \right| \phi \right>
= \left< \phi_{i}^{a} \right| \bar{H}_N(t) \left| \phi \right>,
\end{equation}
with boundary conditions $t_i^a(0)=0$,
where the matrix elements
are obtained from
expressions with matrix products up to third order in the CC amplitudes,
\begin{equation}
\label{eqn:matel2}
\begin{split}
\left< \phi_{i}^{a} \right| \bar{H}_N(t) \left| \phi \right> &=
 f_{ai} + \sum_b f_{ab} t_i^b - \sum_j f_{ji} t_j^a\\
& + \sum_{jb} v_{aj}^{ib} t_j^b - \sum_{jb} f_{jb} t_i^b t_j^a \\
&- \sum_{jkb} v_{ib}^{jk} t_j^a t_k^b + \sum_{jbc} v_{aj}^{bc} t_i^b t_j^c\\
&- \sum_{jkbd} v_{jk}^{bd} t_i^b t_j^a t_k^d.
\end{split}
\end{equation}
It is important to note that these matrix elements are analogous
to those obtained with the standard CCSD approximation to the ground state
when only singles are included (i.e., $T_2=0$).

We can now
introduce the explicit forms of the $f_{pq}$ elements to make the results in
Eq.\ (\ref{eqn:matel1}) and (\ref{eqn:matel2}) more explicit.
First the exact form of the cumulant in Eq.\ (35) is defined by the matrix
element $\left< \phi \left| \bar{H}_N(t) \right| \phi \right>$ and
given by the compact expression with terms
linear (L) and non-linear (NL) in the amplitudes $t_i^a$ and 
first order in the couplings $v_{ij}^{ab}$,
\begin{equation}
\label{eqn:matel3}
\left< \phi \left| \bar{H}_N(t) \right| \phi \right> =
-\sum_{ia} v_{ci}^{ca} t_i^a +
\frac{1}{2} \sum_{ijab} v_{ij}^{ab} t_i^a t_j^b.
\end{equation}
The linear (L) term corresponds to the coupling between the core-hole and the
particle-hole excitation $i \rightarrow a$, while the quadratic terms (NL) represent valence polarization effects that characterize the screening of the
the core-hole.
Similarly the matrix elements for the EOM of the CC amplitudes are
\begin{equation}
\label{eqn:matel4}
\begin{split}
\left< \phi_{i}^{a} \right| \bar{H}_N(t) \left| \phi \right> 
&= -v_{ac}^{ic} + \left( \epsilon_a - \epsilon_i \right) t_i^a \\
&+ \sum_{j} v_{jc}^{ic} t_j^a - \sum_{b} v_{ac}^{bc} t_i^b
+ \sum_{jb} v_{ja}^{bi} t_j^b \\
&+ \sum_{jb} v_{jc}^{bc} t_i^b t_j^a + \sum_{jbd} v_{aj}^{bd} t_i^b t_j^d
- \sum_{jkb} v_{jk}^{ib} t_j^a t_k^b \\
&- \sum_{jkbd} v_{jk}^{bd} t_i^b t_j^a t_k^d.
\end{split}
\end{equation}
Note that the exponential form of Eq.\ (\ref{eqn:cum_form})
is identical to that in our original paper.\cite{RVKNKP}
However,
the differential
equations for the CC coefficients have terms up to third order.
Note also that if one keeps only the first two terms on the RHS of
Eq.\ (\ref{eqn:matel4}),
the cumulant becomes that for the 2nd-order self energy
in Eq.\ (\ref{eqn:cum_t_2ose}). As discussed below, however, the non-linear term turn out to be crucial for accurate calculations.

The result for this more general EOM-CC cumulant can also be represented in
Landau form with a cumulant kernel $\beta(\omega)$ given by 
\begin{equation}
\label{eq:beta}
\beta(\omega) =\frac{1}{\pi}{\rm Re}\, \int_0^{\infty} dt\, e^{-i\omega t}
\frac{d}{dt}\langle\phi|\bar{H}_N(t)|\phi\rangle.
\end{equation}
In contrast to the expression in terms of the 2nd-order self-energy,
the general
EOM-CC kernel $\beta(\omega)$ implicitly contains non-linear terms that
give corrections to 2nd-order approximations in the SE2 or GW formulations.
As a consequence $\beta(\omega)$ is no longer guaranteed to be positive
definite, and similarly the spectral function no longer has
multiple-satellites, consistent with the the particle-hole nature of the
excitations.\cite{marilena20}
Finally we also note that a one-particle form of Eq.\ (\ref{eqn:matel2}) for the CC amplitudes can lead to simpler methods based on alternative one-body, effective Hamiltonians (see SI Sec.\ I.\ E.).

\section{Results}
\label{sec:res}

\subsection{Computational Details}

In this section we illustrate our approach for calculations of the
spectral function for the ten electron ($10e$) series systems: CH$_4$, NH$_3$, H$_2$O, HF and Ne, using different levels of approximation
to the EOM-CCS approach and cumulant, as well as different basis sets.
Our calculations use experimental geometries\cite{cccbdb} for all the molecules:
$r_\mathrm{CH}=$ 1.087\AA,  $r_\mathrm{NH}=$ 1.012\AA, $a_\mathrm{HNH}=$
106.67\degree, $r_\mathrm{OH}=$ 0.958\AA, $a_\mathrm{HOH}=$ 104.48\degree, 
$r_\mathrm{FH}=$ 0.917\AA.  To assess the effect of basis set quality,
calculations were performed with the DZVP,\cite{pyquante} cc-pVDZ,\cite{ccpvdz}
and aug-cc-pVDZ\cite{augccpvdz} basis sets.
The parameters for the KT, DSE2, CSE2 and EOM-CC methods, the molecular orbital energies $\epsilon_p$ and integrals $v_{pq}^{rs}$,
were computed using PyQuante.\cite{pyquante} To improve the efficiency of these methods (in particular in the case of the EOM-CC method), only integrals greater than 1$\times$10$^{-4}$ au were used, i.e., the integrals were screened after the SCF was properly converged. As shown in SI Sec.\ II.\ A., this approximation greatly reduces the amount of data that needs to be handled and results in faster simulations (by an order of magnitude in some cases), with little effect on the calculated spectral functions. The integrals were not screened for the GFCCSD and GFCC-i(2,3) calculations.
The time integration of the EOM in
Eq.\ (\ref{eq:rt_eom_ccs}) was performed using a 4th-order
Adams-Moulton linear multistep method.\cite{quarteroni2002} To obtain the sharp,
broad energy range and smooth spectral functions shown below, the integration
used a time step of 0.025 au ($\sim$0.6 as) with a total simulation time of 600
au ($\sim$14.5 fs).

\subsection{Levels of Approximation}

The results presented below use four levels of approximation for the EOM-CCS
calculations from Eq.\ (\ref{eqn:cum_form})-(\ref{eqn:matel4}) beyond the
independent particle approximation (Koopmans' theorem):
\begin{enumerate}
\addtocounter{enumi}{-1}

\item Second-order approximation obtained by keeping only the first two terms on
the RHS of Eq.\ (\ref{eqn:matel4}); this yields a cumulant GF identical to that
obtained with the 2nd-order self energy (CSE2).

\item Core approximation obtained by keeping the terms
in 0 plus the dominant corrections to
the 2nd-order approximation, i.e., the first four sums in Eq.\
(\ref{eqn:matel4}); this  includes all linear valence-valence sums plus the
quadratic term from excitations coupled to the core-hole.

\item Quadratic approximation which includes terms in 1 plus the fifth and sixth
sums in Eq.\ (\ref{eqn:matel4}) corresponding to quadratic valence-valence
terms; these new terms give corrections that shift the excitation energies
closer to the QP peak.

\item Full T$_1$ approach obtained by keeping all terms in the EOM-CCS
approximation in Eq.\ (\ref{eqn:matel4}), including the third order term
in the CC amplitudes in the last line of Eq.\ (\ref{eqn:matel4}).

\end{enumerate}
Each of this approximations to Eq. (\ref{eqn:matel4}) can be paired with linear
(L) and non-linear (NL) approximations to the cumulant defined in  Eq.\
(\ref{eqn:cum_t_f}), which we label as 0$_L$, 0$_{NL}$, 1$_L$, 1$_{NL}$, etc.

For comparison we also include results for i) the bare energy or
Koopmans' Theorem (KT); ii) the exact solution of the Dyson equation
using the diagonal 2nd-order self energy (DSE2)
\begin{equation}
G_p(\omega) = [1-G^0_p(\omega)\Sigma^{(2)}_{pp}(\omega)]^{-1}G^0_p(\omega);
\end{equation}
and iii) results from the GFCCSD and GFCC-i(2,3) methods 
(using only the cc-pVDZ and aug-cc-pVDZ basis
sets).\cite{PengKowalski2018,doi:10.1063/1.5046529,doi:10.1021/acs.jctc.9b00172}

\subsection{Quasiparticle properties}

Table \ref{tbl:QPBind} presents a comparison between the experimental core
binding energies of the $10e$ systems to those computed with the aug-cc-pVDZ 
basis set.
We also include KT, DSE2, GFCCSD and
GFCC-i(2,3) results. Equivalent tables for the other basis sets are given
in the SI. While KT seriously overestimates the core-binding energies
the GFCC results (including both GFCCSD and GFCC-i(2,3)) give significant
improvements. Notably the inclusion of inner triples in the GFCC-i(2,3)
reduces the MAE from 4.24 eV to 2.83 eV.
The results from the EOM-CC approach with only the linear (L) approximation
for the cumulant are comparable, with slightly larger
MAEs from 4.75 to 5.28 eV.
Surprisingly the DSE2 results for the quasiparticle peak are slightly better,
with absolute discrepancies of up to 2 eV and a MAE slightly over 1
eV over a broad energy regime.

\begin{table*}[t]
\caption{Comparison of the experimental core binding energies (in eV) to those obtained with the aug-cc-pVDZ basis set, using the L and NL approximations to the cumulant and the 1-3 approximations of the EOM-CCS method, and their mean absolute errors (MAE). The results from GFCCSD and GFCC-i(2,3) are obtained from the coupled-cluster Green's function approaches\cite{PengKowalski2016,PengKowalski2018} and improve on the cumulant with only linear terms in the CC coefficients. However, the inclusion of
the non-linear terms in the cumulant significantly improves the agreement
with experiment.
}
\label{tbl:QPBind}
\begin{ruledtabular}
\begin{tabular}{lddddddddddddl}
\tbhr{System} & \tbhc{KT} & \tbhc{DSE2}&\tbhc{GFCCSD} &\tbhc{GFCC-i(2,3)}  & \tbhc{1$_L$} & \tbhc{2$_L$} & \tbhc{3$_L$} && \tbhc{1$_{NL}$} & \tbhc{2$_{NL}$} & \tbhc{3$_{NL}$} & \tbhc{Expt}& \tbhc{Ref} \\
\hline
CH$_4$& 305.18 & 292.24 & 293.34 & 292.69 & 286.35 & 287.31 & 286.89 && 290.020 & 290.62 & 290.36 & 290.703 &[\onlinecite{ch4spf}]\\
NH$_3$& 423.18 & 405.93 & 409.10 & 407.90 & 400.18 & 400.85 & 400.25 && 404.865 & 405.27 & 404.92 & 405.52 &[\onlinecite{nh3spf}]\\
H$_2$O& 559.91 & 538.97 & 544.28 & 542.76 & 534.15 & 534.23 & 533.56 && 539.225 & 539.28 & 538.89 & 539.7 &[\onlinecite{h2ospf}]\\
HF    & 715.89 & 692.29 & 699.39 & 697.48 & 688.91 & 688.40 & 687.81 && 693.710 & 693.40 & 693.03 & 694.2 &[\onlinecite{hfspf}] \\
Ne    & 892.40 & 868.15 & 875.44 & 873.70 & 866.60 & 865.80 & 865.44 && 870.458 & 869.91 & 869.66 & 870.2 &[\onlinecite{nespf}] \\
\hline
MAE   &  19.25 & 1.33 & 4.24 &   2.83 &   4.83 &   4.75 &	5.28 &&   0.51 &    0.37 &    0.69 &	    &			  \\
\end{tabular}
\end{ruledtabular}
\end{table*}

The quality of the EOM-CCS results depends primarily on the inclusion of
non-linear terms in the cumulant rather than level of approximation
used for the CC amplitudes in Eq.\ (\ref{eq:rt_eom_ccs}), so long as
the non-linear terms are included.  The linear (L) approximation to
the cumulant from the first term on the right in  Eq.\
(\ref{eqn:matel3}) consistently underestimates the binding energy by about 3-6
eV. Notably, the introduction of the NL terms 
in  Eq.\ (\ref{eqn:matel3}) reduces the error by an order
of magnitude, thereby bringing the results for the binding energy
in very good agreement with experiment, with mean absolute errors
(MAE) of 0.7 eV or less. Remarkably all three non-linear approximations
to the EOM-CCS equations (i.e., levels 1$_{NL}$-3$_{NL}$),
produce similar errors, with a systematic
underestimation of the experimental results by less than an eV.
However, the full (3$_{NL}$) treatment of the
T$_1$ term does not improve the trend. This suggests that 
terms beyond the CC-singles approximation
in Eqs.\ (\ref{eq-dlnndt}) and (\ref{eq-dtdt}) are desirable in an effort
to achieve higher accuracy.
Nevertheless, our EOM-CC cumulant results demonstrate
that even the simplest
$T_1$ approximation is capable of recovering most of the relaxation energy
required to reproduce experimental quasiparticle binding energies to within
an eV. This result is consistent with the typically very good
relaxation energies obtained with $\Delta$SCF
approaches. These can be cast in terms of $e^{T_1}$ rotations of the ground
state orbitals, as demonstrated by Thouless' theorem. Finally, for the EOM-CCS
method all three basis sets used here yield about the same MAE,
showing that the quasiparticle energy can be computed to within 1 eV even
with rather modest basis sets. 

Table \ref{tbl:QPZ} shows a comparison between the quasiparticle
strengths (i.e., the renormalization constants) for
the $10e$ systems computed with the best basis set (aug-cc-pVDZ) using the L and
NL approximations to the cumulant and the 1-3 approximations of the
EOM-CCS method. The two approximations to the cumulant show similar trends, with the strength increasing almost systematically in the CH$_4$-Ne series. The DSE2 results do not show this trend, and are systematically higher than the
those for the EOM-CCS. Thus the inclusion of the NL term in
the cumulant has the effect of transferring intensity from the satellites back into the quasiparticle peak.

\begin{table*}[t]
\caption{Comparison of the quasiparticle strengths obtained with the aug-cc-pVDZ basis set, using the L and NL approximations to the cumulant and the 1-3 approximations of the EOM-CCS method. Note that the non-linear terms in
the cumulant significantly increase the quasiparticle strength, while 
the 2nd-order Green's function approximation is even larger. The
quadratic (2) and cubic terms (3) in the EOM have only minor effects.
}
\label{tbl:QPZ}
\begin{ruledtabular}
\begin{tabular}{ldddddddddddd}
\tbhr{System} & \tbhc{DSE2}&\tbhc{GFCCSD} &\tbhc{GFCC-i(2,3)}  & \tbhc{1$_{L}$} & \tbhc{2$_{L}$} & \tbhc{3$_{L}$} && \tbhc{1$_{NL}$} & \tbhc{2$_{NL}$} & \tbhc{3$_{NL}$} &\\
\hline
CH$_4$&0.80&0.80     &  0.76&  0.59&  0.63&  0.60&&  0.70&  0.72&  0.70&\\
NH$_3$&0.77&0.81     &  0.75&  0.60&  0.62&  0.59&&  0.71&  0.72&  0.70&\\
H$_2$O&0.76&0.82     &  0.77&  0.63&  0.63&  0.60&&  0.73&  0.73&  0.71&\\
HF    &0.77&0.84     &  0.78&  0.69&  0.67&  0.65&&  0.76&  0.75&  0.74&\\
Ne    &0.80&0.87     &  0.81&  0.76&  0.73&  0.72&&  0.81&  0.79&  0.78&\\
\end{tabular}
\end{ruledtabular}
\end{table*}

\subsection{Satellite properties}

\begin{figure}[t]
\includegraphics[scale=0.40,clip]{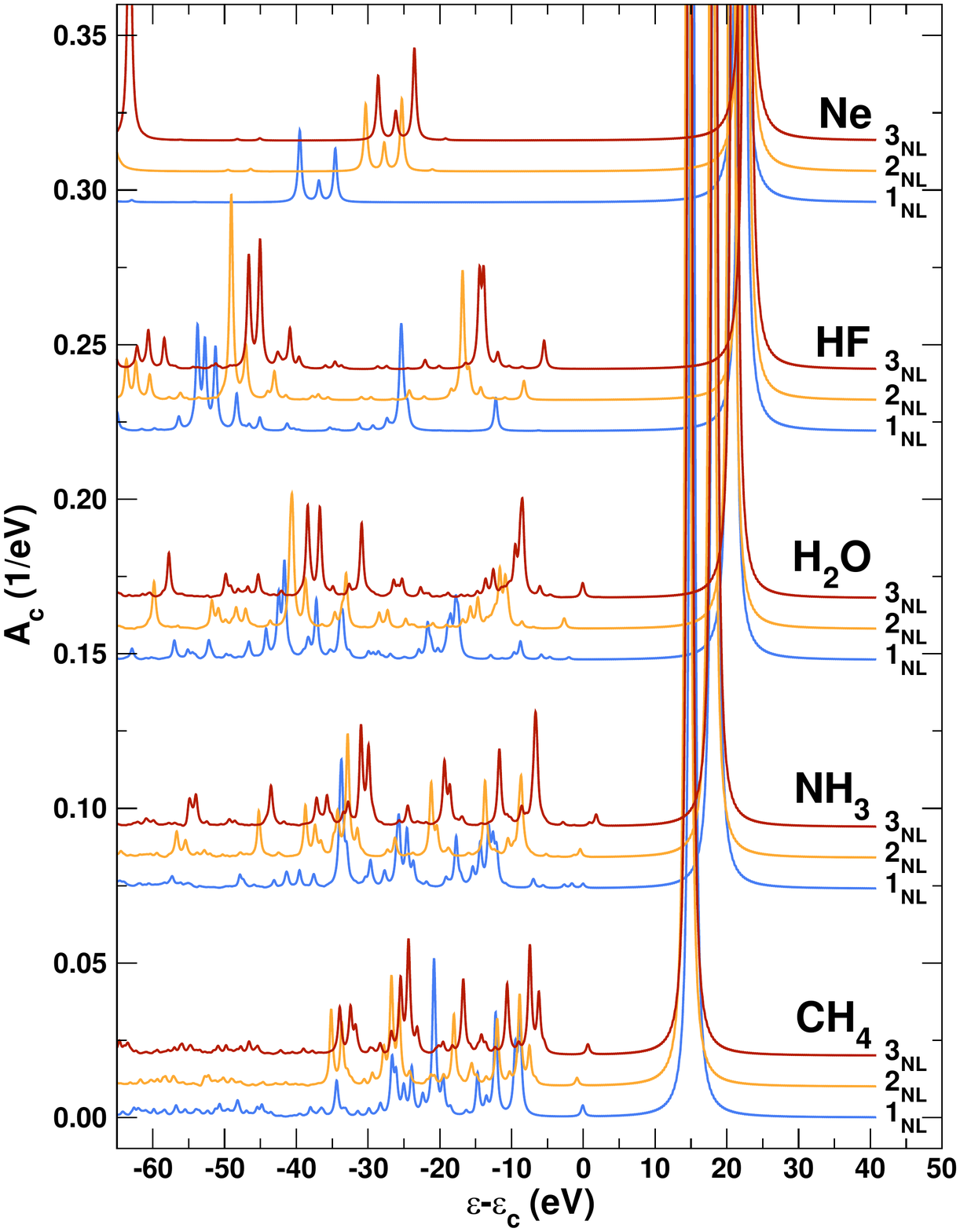}
\caption{\label{fig:10eac}
Core spectral function $A_c$ of the $10e$ systems computed with the aug-cc-pVDZ basis set and the full NL cumulant form, as a function of EOM-CCS approximation 1$_{NL}$ (blue), 2$_{NL}$ (orange), and 3$_{NL}$ (red).
}
\end{figure}
\begin{figure}[t]
\includegraphics[scale=0.36,
trim=1.0cm 1.5cm 0.0cm 2.0cm,clip]{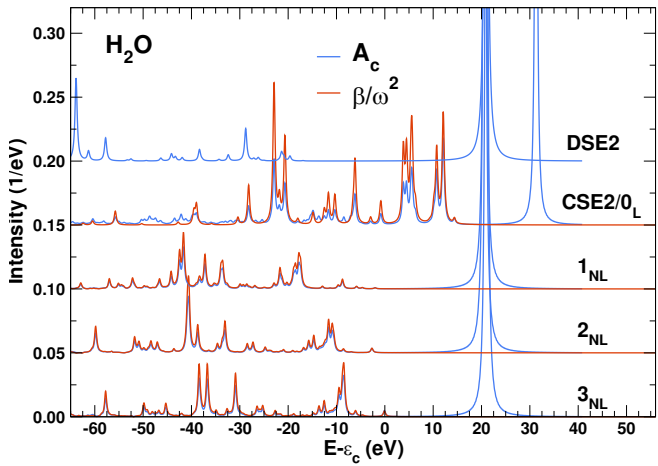}
\caption{\label{fig:h2o_beta}
Comparison of the core spectral function $A_c$ (blue) and the cumulant kernel $\beta(\omega)/\omega^2$ (red) for H$_2$O, computed with the aug-cc-pVDZ basis set and the full NL cumulant form, as a function of approximation (1$_{NL}$-3$_{NL}$) of the EOM-CCS method. For comparison we also include results obtained with the DSE2 and CSE2 methods, the latter being equivalent to the 0$_L$ approximation.
}
\end{figure}

The spectral function $A_c(\omega)$, which
characterizes the excitation spectrum for a given level, 
is defined as
\begin{equation}
A_{c}(\omega) = -\frac{1}{\pi} {\rm Im}\, G_{c}(\omega).
\end{equation}
This spectrum characterizes the intrinsic, inelastic losses in the
system as measured, e.g., in XPS.
Fig.\ \ref{fig:10eac} shows a comparison of $A_c(\omega)$ for the
$10e$ systems computed with the aug-cc-pVDZ basis set and the full NL cumulant
form, as a function of EOM-CCS approximation. Similar figures for the other
basis sets and for the linear approximation to the cumulant are given in the
SI.
To make the comparison between different systems more clear, the spectral
functions in Fig.\ \ref{fig:10eac} are plotted with respect to the bare
core-hole energy $\epsilon_c$ (which corresponds to $E^{B}_{KT} = -\epsilon_c$,
where $E^{B}_{KT}$ is the Koopmans' Theorem binding energy in Table
\ref{tbl:QPBind}). With this reference, the position of the quasiparticle peak
corresponds to the relaxation energy
\begin{equation}
\Delta=\epsilon_B-|\epsilon_c| = \int d\omega \frac{\beta(\omega)}{\omega},
\end{equation}
as discussed in our original treatment.\cite{RVKNKP} Given that the relaxation
is inversely proportional to the mean core-valence interaction potential, which
for these systems decreases from CH$_4$ to Ne, we see a blue shift trend in the
quasiparticle position.

\begin{figure}[t]
\includegraphics[scale=0.28,
trim=0.0cm 1.9cm 0.0cm 1.0cm,clip]{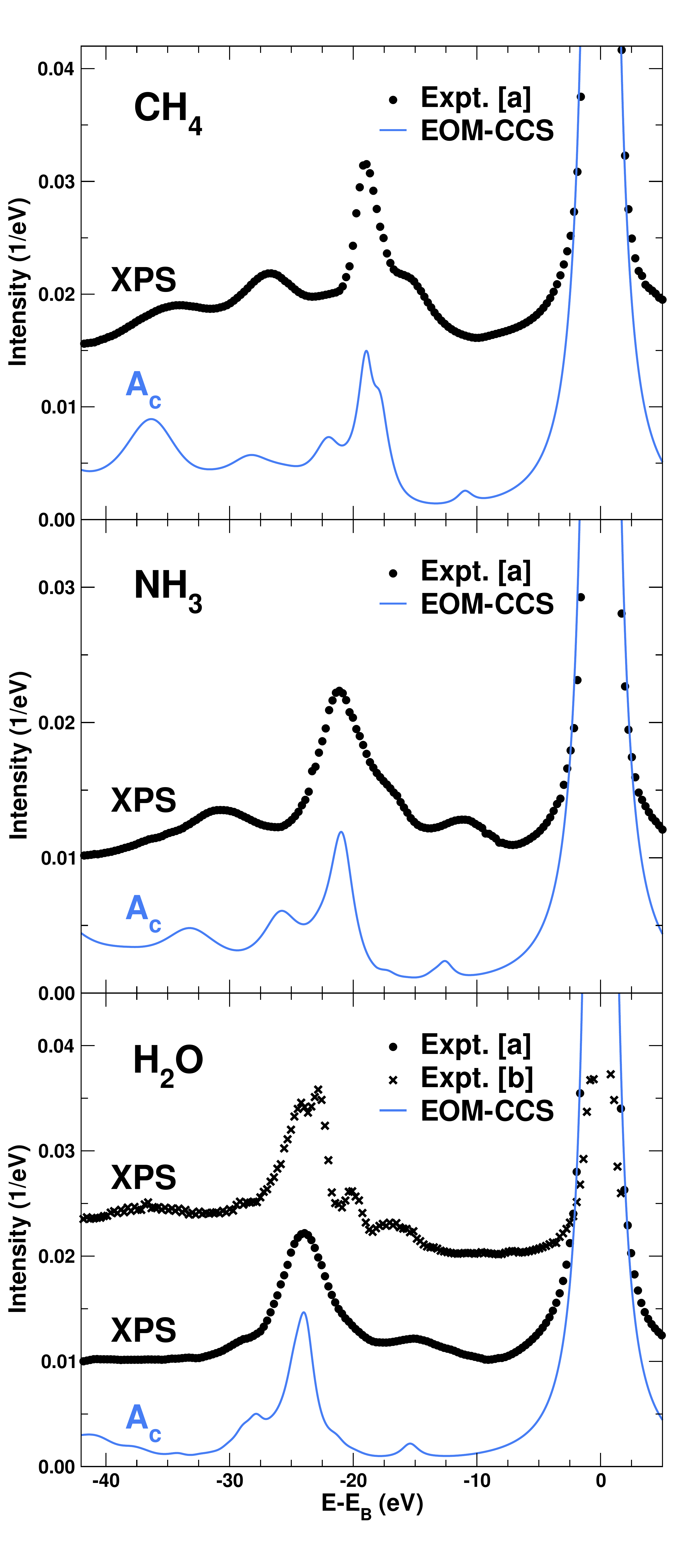}
\caption{\label{fig:A_c_expt}
Comparison between the experimental XPS (Expt.: [a] Ref. \onlinecite{doi:10.1063/1.439722}, [b] Ref. \onlinecite{SANKARI200651}) and the aug-cc-pVDZ (3$_{NL}$) EOM-CCS core spectral functions of CH$_4$ (top), NH$_3$ (middle), and H$_2$O (bottom) as a function of energy. The EOM-CCS results include scissors corrections of 5.7, 3.9 and 3.2 eV, respectively, to close the QP-satellite gap, and have been broadened roughly to match experiment.
}
\end{figure}

Fig.\ \ref{fig:10eac} also shows the satellite region near the quasiparticle
peak. We find that the main effect of EOM-CCS approximation including non-linear
corrections is a narrowing of the quasiparticle-satellite gap improving the
agreement with experiment. Given that the quasiparticle position is nearly
constant with respect to the level of approximation, most of the gap closing
arises from an increase in the satellite energies. For most of the studied
systems, we find that the satellite structure obtained with the quadratic
approximation (2$_{NL}$) is similar to that for the full method (3$_{NL}$) up to
the overall shift (see SI Fig. 6). The linear approximation shows a similar
satellite weight distribution, but with some differences in the position of the
particular features.

Fig. \ref{fig:h2o_beta} shows a comparison between the cumulant kernel
$\beta(\omega)/\omega^2$ and the satellite region of $A_c(\omega)$ for H$_2$O.
As expected, the peaks in $\beta(\omega)$ for the EOM-CC cumulant
correspond well to the inelastic losses in $A_c(\omega)$.
Note however, that neither the DSE2 nor the CSE2 approximations are adequate to
describe the satellite spectra of small molecules.  The CSE2 significantly
overestimates the quasiparticle relaxation energy, while the DSE2 gives a
poor representation of the spectral function.
As discussed in Sec. \ref{sec:rt_core_gf}, given that the CSE2 and the
minimal (0$_L$) approximation to the EOM-CC method
are equivalent, this error 
is corrected by the remaining terms in the EOM-CCS equations which bring
the quasiparticle into good agreement with the DSE2 results and with experiment.

Finally, Fig. \ref{fig:A_c_expt} shows a comparison of the full (3$_{NL}$)
EOM-CCS spectral functions obtained with the aug-cc-pVDZ basis set, to available
experimental XPS results. We focus mainly on the satellite region, and the
EOM-CCS results include scissors corrections to facilitate the comparison to
experiment. We find that the agreement with experiment is quite satisfactory
considering the simplicity of the EOM-CCS approximation. The satellite peaks
arise from shake-up excitations, in which the creation of the core hole is
accompanied by valence-valence excitations, i.e., density fluctuations of
opposite sign.  The need to include scissors corrections indicates that the
$e^{T_1}$ approximation is not sufficient to fully account for the  relaxation
of the satellite transitions due to electron-electron interactions and dynamic
screening of the core-hole. However, the reasonable agreement obtained using a
single scissors shift for each system points to missing dynamic correlation
effects that would likely be accounted for with the inclusion of higher order
cluster operators. For example, comparing GFCCSD and GFCC-i(2,3) results, we did
find that the satellite positions shift significantly towards the main
quasiparticle peaks. For H$_2$O, the shift brought by the triples can be as
large as $\sim$20 eV. However, the inclusion of higher order terms in the time
domain in an effective and economic way is still an open question and is one of
the future directions of research. 

\section{Conclusions}

We have investigated the cumulant representation of the one-particle Green's
function based on a real-time EOM-CC approximation. The logarithmic-derivative
behavior of the EOM-CC yields an explicit exponential cumulant representation of
the GF, with a non-perturbative expression for the cumulant in terms of the
solutions to a set of  coupled, first-order non-linear differential equations.
Within the CC-singles approximation implemented here, the non-linear terms enter
the formulation in two ways. First the expression for the cumulant is quadratic
in the CC amplitudes $t_{ia}(t)$. Second, the EOM for the CC amplitudes have
non-linear terms up to third order. To lowest order, i.e., 2nd-order in the
electron-electron interaction, the retarded cumulant reduces to that defined by
the 2nd-order self energy (SE2). We also found that the time-kernel in the
EOM-CC is directly analogous to that in the static CC equations for the ground
state, thus simplifying the implementation of our approach in current codes. As
a quantitative test, we have applied the EOM-CC cumulant approach to the
core-hole Green's function for a number of small molecular systems. We find that
the EOM-CC cumulant approach with only CC-singles excitations yields accurate
quasiparticle properties such as core-level binding energies, as well as an
approximate treatment of satellite shapes. The observed improvement over other
approaches likely stems from the implicit excitations present in the exponential
form of the cumulant ansatz. Although the 2nd-order approximation gives
reasonable results for the quasiparticle energy with the Dyson equation, the
2nd-order cumulant approximation for the relaxation energy has substantial
errors, and the spectral function is poorly described. This behavior is in
contrast to that observed in condensed matter, where the $GW+C$
approach\cite{sky,PhysRevLett.110.146801,PhysRevB.94.035103} based on a cumulant
from the 2nd-order GW self-energy and the quasi-boson approximation has proved
to be advantageous for a description of satellites in the spectral
function.\cite{Hedin99review,sky} Physically, this difference appears to reflect
the particle-hole character of excitations in molecular systems, which leads to
a sparser satellite structure without multiple-bosonic
excitations.\cite{marilena20} This suggests that the inclusion of non-linear
corrections to the cumulant is generally important for understanding the nature
of electronic excitations in molecular systems.

\medskip

\noindent Data Availability Statement: The data that support the
findings of this study are available from the corresponding author
upon reasonable request.

\begin{acknowledgments}
 This work was supported by the Computational
Chemical Sciences Program of the U.S. Department
of Energy, Office of Science, BES, Chemical Sciences, Geosciences
and Biosciences Division in the Center for Scalable
and Predictive methods for Excitations and Correlated phenomena
(SPEC) at PNNL. 
\end{acknowledgments}

\bibliography{EOM-CC-Cumulant-Approach}
\bibliographystyle{apsrev}

\end{document}


\title{Supplementary Information: Real-time coupled-cluster approach for the cumulant Green's function}
\date{\today}

\author{F. D. Vila}
\author{J. J. Rehr}
\author{J. J. Kas}
\affiliation{Department of Physics, University of Washington, Seattle, WA 98195}
\author{K. Kowalski}
\affiliation{William R. Wiley Environmental Molecular Sciences Laboratory, Battelle, Pacific Northwest National Laboratory, K8-91, P. O. Box 999, Richland, Washington 99352}
\author{B. Peng}
\affiliation{Physical Sciences Division, Pacific Northwest National Laboratory, Richland, WA 99354}

\begin{abstract} 
This Supplementary Information presents results that are too detailed for the main manuscript, including more detailed discussion of the theory derivations in the main manuscript, computational details, results for other basis sets, and more results that are too technical for the main manuscript.
\end{abstract}

\date{\today}

\maketitle

\section{Theory}

\subsection{Retarded Green's function in spin-orbital basis}

The real-space, real-time retarded Green's function is defined as
\begin{equation}
\label{eq:gfrtrs}
\begin{split}
G(x,x') =& ~ G(rt,r't') = \\
&-i \Theta(t-t')\left<0\left| \left\{\psi(rt), \psi^\dagger(r't') \right\} \right| 0 \right>
\end{split}
\end{equation}
where $\left| 0 \right>$ is the ground state of the system, $\psi^\dagger(r't')$
and $\psi(rt)$ are, respectively, the creation and annhilation field operators
at $r't'$ and $rt$, and $\left\{,\right\}$ indicates the anticonmutation
operator. By introducing a basis set of single-particle spin-orbitals
$\{\phi_p(r)\}$, we can express any two-positions, two-times operator $M$ as:
\begin{equation}
\label{eqn:mel1}
M(rt,r't') = \sum_{pq} \phi^{*}_{p}(r) M_{pq}(t,t') \phi_{q}(r')
\end{equation}
where the matrix elements are defined as:
\begin{equation}
\label{eqn:mel2}
M_{pq}(t,t') = \int dr dr' \phi^{*}_{p}(r) M(rt,r't') \phi_{q}(r').
\end{equation}
Inserting these definitions into Eq. (\ref{eq:gfrtrs}) we obtain the Green's function matrix expressed in the spin-orbital basis $\{\phi_p(r)\}$:
\begin{equation}
\label{eqn:gpq}
G_{pq}(t,t') =
-i \Theta(t-t')
\left<0\left| \left\{a_p(t), a_q^\dagger(t') \right\} \right| 0 \right>,
\end{equation}
where the creation and annhilation operators $a_p^\dagger(t)$ and $a_q(t')$ are
associated with the spin-orbitals $\phi_p$ and $\phi_q$ respectively.

\subsection{From the real-space time-domain Dyson equation to the cumulant ansatz in a spin-orbital basis}

The real-space, time-domain form of the Dyson equation
\begin{equation}
\begin{split}
G(x,x') =&~ G^0(x,x') + \\
&\int dx_1 dx_2 G^0(x,x_1) \Sigma(x_1,x_2) G(x_2,x')
\end{split}
\end{equation}
can also be cast in matrix form (after including time translation invariance):
\begin{equation}
\label{eqn:dyson}
\hat{G}(t) = \hat{G}^0(t) + \int dt_1 dt_2 \hat{G}^0(t-t_1)
\hat{\Sigma}(t_1-t_2) \hat{G}(t_2).
\end{equation}
Expanding the matrix form of the cumulant ansatz
\begin{equation}
\label{eqn:cummat}
\hat{G}(t) = \hat{G}^0(t)e^{\hat{C}(t)},
\end{equation}
and Eq. (\ref{eqn:dyson}) to first order
\begin{eqnarray}
\hat{G}(t) &=& \hat{G}^0(t) + \hat{G}^0(t)\hat{C}(t) + ... \\
\hat{G}(t) &=& \hat{G}^0(t) + \int dt_1 dt_2 \hat{G}^0(t-t_1)
\hat{\Sigma}(t_1-t_2) \hat{G}^0(t_2) + ...
\end{eqnarray}
and equating the second terms in the right hand sides we get:
\begin{equation}
\hat{G}^0(t)\hat{C}(t) = \int dt_1 dt_2 \hat{G}^0(t-t_1)
\hat{\Sigma}(t_1-t_2) \hat{G}^0(t_2)
\end{equation}
For a typical matrix element we obtain
\begin{equation}
\sum_r G^0_{pr}(t) C_{rq}(t) = \sum_{rs} \int dt_1 dt_2 G^0_{pr}(t-t_1)
\Sigma_{rs}(t_1-t_2) G^0_{sq}(t_2),
\end{equation}

Shifting the $\omega$ integration variable and using the Fourier transform of
the cumulant we have:
\begin{equation}
C_{pq}(\omega) = i G^0_{pp}(\omega+\epsilon_p)
\Sigma_{pq}(\omega+\epsilon_p) G^0_{qq}(\omega+\epsilon_p).
\end{equation}
Finally, introducing the frequency domain form of
$G^0_{pp}=(\omega-\epsilon_p+i\delta)^{-1}$:
\begin{equation}
\label{eqn:cumfreq}
C_{pq}(\omega) = \frac{i \Sigma_{pq}(\omega+\epsilon_p)}
{(\omega+i\delta)(\omega+\epsilon_p-\epsilon_q+i\delta)}.
\end{equation}
If we approximate $\Sigma_{pq}(\omega) \simeq \Sigma_{pp}(\omega) \delta_{pq}$,
then
\begin{equation}
\label{eqn:cum_diag}
C_{pp}(\omega) = \frac{i \Sigma_{pp}(\omega+\epsilon_p)}
{(\omega+i\delta)^2},
\end{equation}
or, in the time domain
\begin{equation}
\label{eqn:cum_diag_t}
C_{pp}(t) = \int \frac{d\omega}{2\pi} \frac{i\Sigma_{pp}(\omega+\epsilon_p)}
{(\omega+i\delta)^2} e^{-i\omega t}.
\end{equation}
We can now return to the time domain, inserting into Eq.\ (\ref{eqn:cummat}) and
taking into account that the matrix exponential is now diagonal:
\begin{equation}
\label{eqn:g_cum_diag}
G_{pp}(t) = i \Theta(t) e^{-i \epsilon_p t + C_{pp}(t)},
\end{equation}
which is the standard form of the diagonal cumulant.

\subsection{Coupled Cluster Green's function in time}
\label{sec:ccg_t}

In this section we derive a compact form for the full Coupled Cluster Green's function that can be used for further derivation of time-domain approximations. Starting with Eq.\ (17)
in Ref. \onlinecite{PhysRevA.94.062512} (from now on referred as ``PK'') we have
\begin{equation}
\label{eq:GPK1}
G^R_{pq} (\omega) = \left< \Phi \left| (1+\Lambda) e^{-T} a^{\dagger}_q
\left( \omega + (H-E_0) -i\delta \right)^{-1}
a_p e^{T} \right| \Phi \right>
\end{equation}
and inserting the $I=e^{-T}e^{T}$ we get
\begin{equation}
\label{eq:GPK2}
G^R_{pq} (\omega) = \left< \Phi \left| (1+\Lambda) \bar{a^{\dagger}_q}
\left( \omega + \bar{H}_N \right)^{-1}
\bar{a}_p \right| \Phi \right>
\end{equation}
where $\bar{O} = e^{-T}Oe^{T}$ is the similarity transformed form of the $O$
operator, $H_N$ is the normal ordered hamiltonian, and using the
Baker-Campbell-Hausdorff (BCH) relation we have that
\begin{equation}
\label{eq:apbar}
\bar{a}_p = a_p + [a_p,T],
\end{equation}
\begin{equation}
\label{eq:aqbar}
\bar{a^{\dagger}_q} = a^{\dagger}_q + [a^{\dagger}_q,T].
\end{equation}
For simplicity, we now make the convergence factor $-i\delta$ implicit in the energy $\omega$.

Following PK, we introduce the $X_p(\omega)$, $Z_q(\omega)$, and $W_q(\omega)$
operators, which are solutions to the following equations:
\begin{equation}
X_{p}(\omega) \left| \Phi \right> = 
\left( \omega + \bar{H}_N \right)^{-1}
\bar{a}_p \left| \Phi \right>,
\end{equation}
\begin{equation}
\left< \Phi \right| Z_{q}(\omega) = 
\left< \Phi \right| (1+\Lambda) \bar{a^{\dagger}_q}
\left( \omega + \bar{H}_N \right)^{-1},
\end{equation}
and
\begin{equation}
\left< \Phi \right| (1+\Lambda) W_{q}(\omega) = 
\left< \Phi \right| (1+\Lambda) \bar{a^{\dagger}_q}
\left( \omega + \bar{H}_N \right)^{-1}
\end{equation}
These operators have the following expansions in the $N-1$ Fock space:
\begin{equation}
\label{eq:Xomega}
\begin{split}
X_{p}(\omega) =& \sum_i x^i(\omega)_p a_i +\\
&\frac{1}{2!}\sum_{ij,a} x^{ij}_a(\omega)_p a^{\dagger}_a a_j a_i + ...
\end{split}
\end{equation}
\begin{equation}
\label{eq:Zomega}
\begin{split}
Z_{q}(\omega) =& \sum_i z_i(\omega)_q a^{\dagger}_i +\\
&\frac{1}{2!}\sum_{ij,a} z_{ij}^a(\omega)_p a^{\dagger}_i a^{\dagger}_j a_a + ...
\end{split}
\end{equation}
and
\begin{equation}
\label{eq:Womega}
\begin{split}
W_{q}(\omega) =& \sum_i w_i(\omega)_q a^{\dagger}_i +\\
&\frac{1}{2!}\sum_{ij,a} w_{ij}^a(\omega)_p a^{\dagger}_i a^{\dagger}_j a_a + ...
\end{split}
\end{equation}

Eqs. (\ref{eq:GPK1})-(\ref{eq:Womega}) are a summary of the formulation in PK. Now, using
\begin{equation}
\mathrm{IFT}\left[\left( \omega + \bar{H}_N -i\delta \right)^{-1} \right] =
i \Theta(-t) e^{i \bar{H}_N t}
\end{equation}
where $\mathrm{IFT}$ is the inverse Fourier transform, we can write
$G^R_{pq} (\omega)$ in the time-domain as:
\begin{equation}
G^R_{pq}(t) = i \left< \Phi \left| (1+\Lambda) \bar{a^{\dagger}_q}
e^{i \bar{H}_N t}
\bar{a}_p \right| \Phi \right>
\end{equation}
where from now on we assume that $t<0$ to remove all the $\Theta(-t)$ functions.
We can also convert the equations defining the $X_p(\omega)$, $Z_q(\omega)$,
and $W_q(\omega)$ operators as:
\begin{equation}
\label{eq:xpt}
X_{p}(t) \left| \Phi \right> = 
i e^{i \bar{H}_N t}
\bar{a}_p \left| \Phi \right>,
\end{equation}
\begin{equation}
\label{eq:zqt}
\left< \Phi \right| Z_{q}(t) = 
i \left< \Phi \right| (1+\Lambda) \bar{a^{\dagger}_q}
e^{i \bar{H}_N t}
\end{equation}
and
\begin{equation}
\label{eq:wqt}
\left< \Phi \right| (1+\Lambda)W_{q}(t) = 
i \left< \Phi \right| (1+\Lambda) \bar{a^{\dagger}_q}
e^{i \bar{H}_N t}
\end{equation}
where the $X_p(t)$, $Z_q(t)$, and $W_q(t)$ are defined simply by the inverse
Fourier transform of their coefficients in their excitation expansions (Eqs. (\ref{eq:Xomega})-(\ref{eq:Womega})).

Based on the properties of the $\bar{H}_N$, i.e., where one assumes that the CC
equations $Q\bar{H}_N|\Phi\rangle = 0$ are satisfied, where $Q$ denotes the
projection operator onto the space spanned by excitations with respect to
$|\Phi\rangle$ Slater determinants, we can prove that, in full analogy with
the frequency representation, the $X_p(t)$ operator can be expressed in terms
of connected diagrams only. To this end we expand $X_p(t)$ in term of powers of
the $\bar{H}_N$ operator
\begin{eqnarray}
X_p(t)|\Phi\rangle &=& ie^{i\bar{H}_Nt} \bar{a_p}|\Phi\rangle \;, \label{step1} \\
&=& i \lbrace \sum_{n=0}^{\infty} \frac{1}{n!} (i)^n t^n (\bar{H}_N)^n \bar{a_p}\rbrace |\Phi\rangle \label{step4} 
\end{eqnarray}

\begin{figure}
\includegraphics[trim= 0.0in 0.5in 0.0in 0.0in, width=0.65\textwidth]{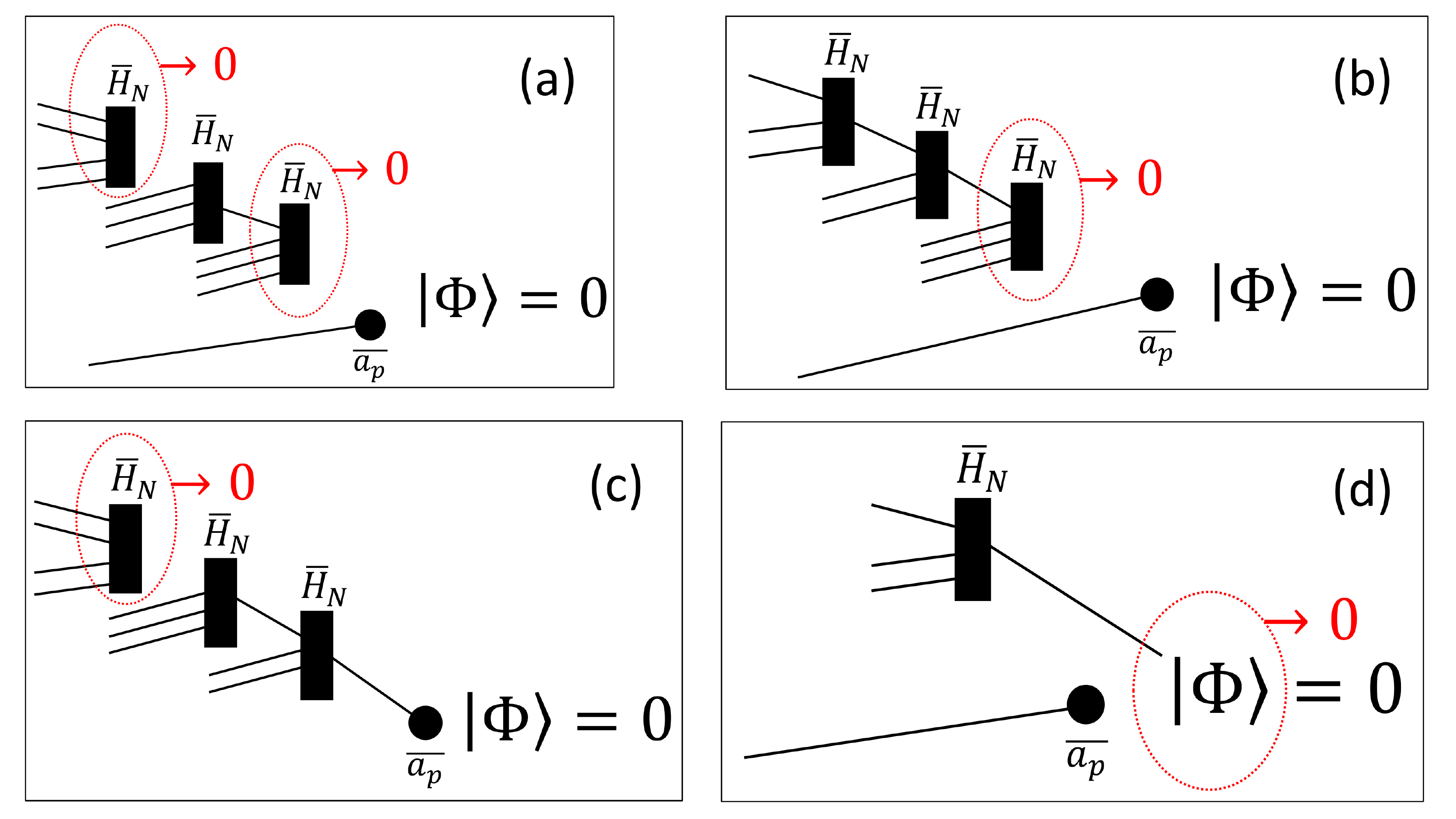}
\caption{Typical examples of diagrams not contributing to the general term in Eq. (\ref{gterm}).}
\label{fig2_conn}
\end{figure}

\begin{figure}
\includegraphics[trim= 0.0in 2.0in 0.0in 1.0in, width=0.45\textwidth]{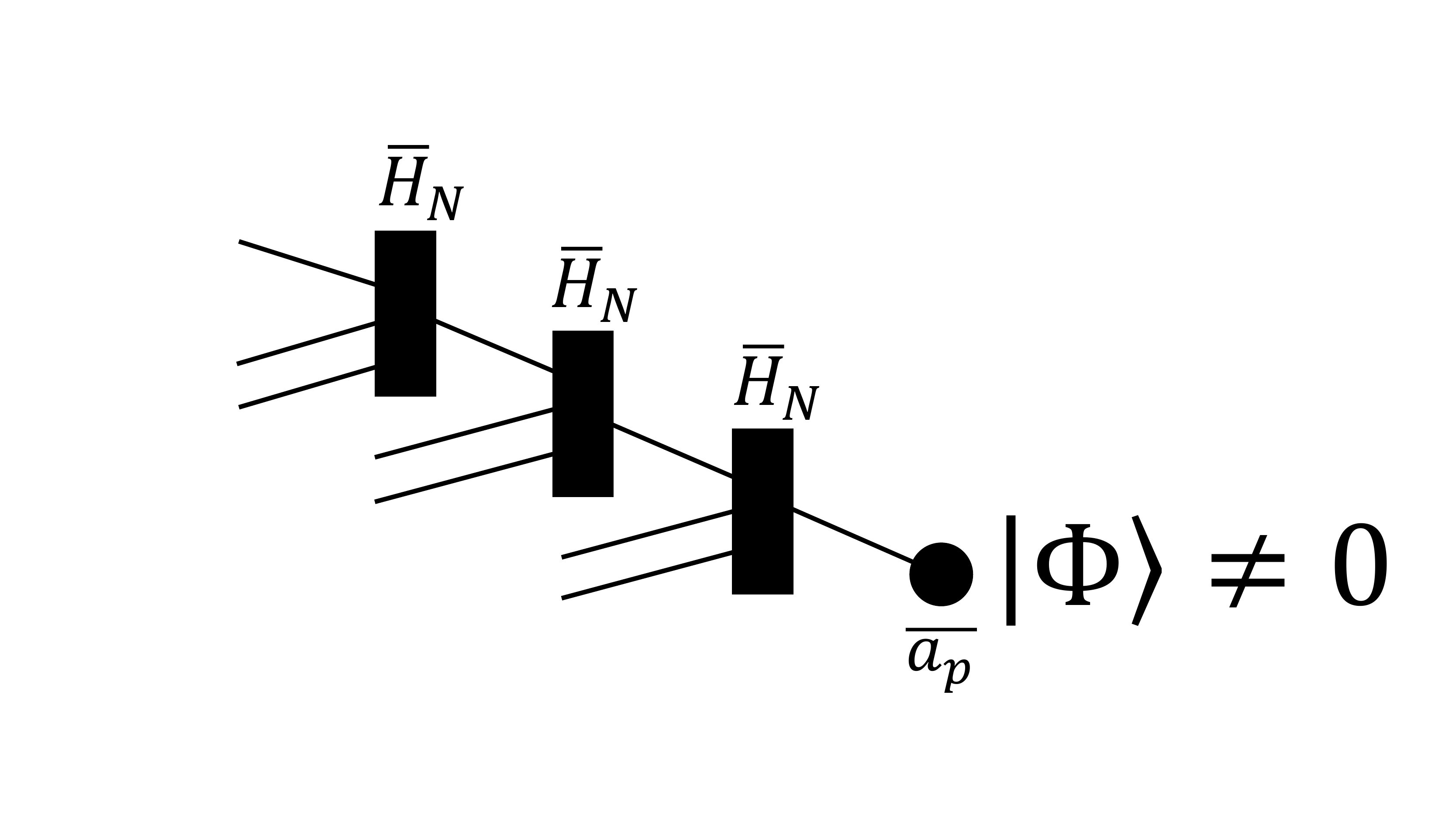}
\caption{Example of a connected diagram that contributes to the general term in Eq. (\ref{gterm}).}
\label{fig3_conn}
\end{figure}

Now we assume that we are dealing with the exact CC theory. This assumption
plays a crucial role in proving the connected character of $X_p(t)$ since all
approximate approaches can be build using connected properties of the exact 
formulation. Using Wick's theorem for the particle-hole formalism to analize a general term in the expansion in Eq. (\ref{step4}):
\begin{equation}
(\bar{H}_N)^n \bar{a_p} |\Phi\rangle,
\label{gterm}
\end{equation}
we can identify several classes of diagrams contributing to $(\bar{H}_N)^n
\bar{a_p} |\Phi\rangle$ (see Figs. (\ref{fig2_conn}) and (\ref{fig3_conn})). It can be easily verified that all disconnected diagrams (typical examples of
these are shown in Fig. (\ref{fig2_conn}a-\ref{fig2_conn}d)) disappear due
to the existence of vertices representing projections of CC equations (i.e., $\bar{H}_N$ matrix elements with all particle-hole creation lines,
that is, all ``legs'', located to the left of the corresponding matrix element or diagrammatic vertex, see Fig. (\ref{fig2_conn}a-\ref{fig2_conn}c)) or due to the existence of the uncontracted particle-hole line (lines) that annihilates the reference function $|\Phi\rangle$ as shown in Fig. (\ref{fig2_conn}d). Consequently, the only diagrams contributing to $X_p(t)$ are connected diagrams
(Fig.(\ref{fig3_conn})) which can be symbolically denoted as 
\begin{equation}
X_p(t)|\Phi\rangle = i
\lbrace e^{i\bar{H}_Nt }\bar{a_p} \rbrace_C |\Phi\rangle\;,
\label{connxp}
\end{equation}
where the subscript ``C'' designates connected part of a given operator expression. 
%

Using the equations above for $X_p(t)$, $Z_q(0^-)$, and $W_q(0^-)$, we can re-write $G^R_{pq}(t)$ as:
\begin{equation}
\label{eq:GZX}
\begin{split}
G^R_{pq}(t)=& \left< \Phi \left| (1+\Lambda)
              \bar{a^{\dagger}_q} X_p(t) \right| \Phi \right> \\
           =& -i \left< \Phi \left|
                Z_q(0^-) X_p(t) \right| \Phi \right> \\
           =& -i \left< \Phi \left| (1+\Lambda)
                W_q(0^-) X_p(t) \right| \Phi \right>
\end{split}
\end{equation}
The first equality is just the time-domain version of Eq. (27) in PK. Note that the $Z_q$ and $W_q$ operators are computed as the limit $ t\rightarrow 0^-$ to avoid any issues with the poorly defined nature of
$\Theta(-t)$ at 0.

We now proceed to expand the second equality in Eq. (\ref{eq:GZX}). We choose to expand this form of $G^R_{pq}(t)$ because it does not include the $\Lambda$ terms (which are implicit in the definition of $Z_{q}(0^-)$) and is thus simpler. Inserting the time dependent definitions of $X_p(t)$ and $Z_q(0^-)$
into $G^R_{pq}(t)$ we have
\begin{multline}
\label{eq:gr_t_sum1}
iG^R_{pq}(t)=
\left< \Phi \right|
\bigl(\sum_i z_i(0^-)_q \{a^{\dagger}_i\} + 
\frac{1}{2!}\sum_{ij,a} z_{ij}^a(0^-)_p \{a^{\dagger}_i a^{\dagger}_j a_a\} + ...\bigr)\\
\bigl(\sum_k x^k(t)_p \{a_k\} +
\frac{1}{2!}\sum_{kl,b} x^{kl}_b(t)_p \{a^{\dagger}_b a_l a_k\} + ...\bigr)
\left| \Phi \right>
\end{multline}
where, to simplify the rest of the calculations, we also introduced the normal
ordered forms of the operators, indicated using Bartlett's style notation
$\{...\}$ rather than the PK one with $N[...]$. Before going any further it is helpful to analyze the general form of a generic matrix element for the different products:
\begin{equation}
\left< \Phi \right|
\{\underbrace{a^{\dagger}_i a^{\dagger}_j ... a_b a_a }_{n~\mathrm{ops}}\}
\{\underbrace{a^{\dagger}_c a^{\dagger}_d ... a_l a_k }_{m~\mathrm{ops}}\}
\left| \Phi \right> =
\sum_\mathrm{FC}
\left< \Phi \right|
\wick{
\{\c3 a^{\dagger}_i \c4 a^{\dagger}_j ... \c1 a_b \c2 a_a
  \c1 a^{\dagger}_c \c2 a^{\dagger}_d ... \c3 a_l \c4 a_k \}
}
\left| \Phi \right>
\end{equation}
where, according to the generalized Wick's theorem (GWT), we compute the Fermi
vacuum expectation value of the product of two normal ordered operator sets by
summing over all possible full contractions (FC). However, for $n \neq m$, the
are no possible full contractions between the two operator sets, thus, the
``cross'' terms in Eq.\ (\ref{eq:gr_t_sum1}) are all zero. For example, we are
only left with terms of the form:
\begin{equation}
\left< \Phi \right|
\{a^{\dagger}_i \} \{a_k \}
\left| \Phi \right> =
\left< \Phi \right|
\wick{
\{\c1 a^{\dagger}_i \c1 a_k \}
}
\left| \Phi \right> = \delta_{ik}
\end{equation}
for 0 excitation order (\textit{i.e.} order 1 in PK),
\begin{multline}
\left< \Phi \right|
\{a^{\dagger}_i a^{\dagger}_j a_a \}
\{a^{\dagger}_b a_l a_k \}
\left| \Phi \right> =
\left< \Phi \right|
\wick{
\{ \c2 a^{\dagger}_i \c3 a^{\dagger}_j \c1 a_a
   \c1 a^{\dagger}_b \c2 a_l \c3 a_k \}
}
\left| \Phi \right>
+
\left< \Phi \right|
\wick{
\{ \c3 a^{\dagger}_i \c2 a^{\dagger}_j \c1 a_a
   \c1 a^{\dagger}_b \c2 a_l \c3 a_k \}
}
\left| \Phi \right> = \\
- \delta_{il}\delta_{jk}\delta_{ab} + \delta_{ik}\delta_{jk}\delta_{ab}
\end{multline}
for 1 excitation order (\textit{i.e.} order 2 in PK), etc. The pattern is fairly
clear: each order produces $h! \times p!$, where $h$ and $p$ are the number of
hole and particle operators in the products. Thus PK order 1 produces one term,
order 2 produces 2, order 3 would produce 12, etc. Thus we can write:
\begin{equation}
iG^R_{pq}(t)=
\sum_i z_i(0^-)_q x^i(t)_p +
\frac{1}{4}\sum_{ij,a} z_{ij}^a(0^-)_p x^{ij}_a(t)_p - 
\frac{1}{4}\sum_{ij,a} z_{ij}^a(0^-)_p x^{ji}_a(t)_p + ...
\end{equation}
This sum can be generalized by recognizing that the $x^{ij}_a(t)_p$ coefficients
are antisymmetric with respect to index swaps, and that $x^{ii}_a(t)_p=0$, thus
\begin{equation}
\begin{split}
iG^R_{pq}(t)=&\\
&\sum_i z_i(0^-)_q x^i(t)_p + \\
+\frac{1}{2}&\sum_{ij,a}   z_{ij}^a(0^-)_p           x^{ij}_a(t)_p + \\ 
+\frac{1}{6}&\sum_{ijk,ab} z_{ijk}^{ab}(0^-)_p x^{ijk}_{ab}(t)_p + ...
\end{split}
\end{equation}

\subsection{Perturbation theory of $G^R_{pq}(t)$}

In this section we demonstrate that the full form of the Coupled Cluster Green's function can be reduced to a cumulant form when approximated as a perturbations series. Although this analysis is valid for any CC level (CCD, CCSD, CCSDT, etc.) with HF orbitals and Moller Plesset MBPT, here for simplicity we limit the CC expansion to doubles ($T=T_2=\frac{1}{4}\sum_{ijab}t_{ij}^{ab} a^{\dagger}_a a^{\dagger}_b a_j a_i$, $\Lambda=\Lambda_2=\frac{1}{4}\sum_{ijab}\lambda_{ij}^{ab} a^{\dagger}_i a^{\dagger}_j a_b a_a$) and expand the $\bar{a}_p$ and $\bar{a^{\dagger}_q}$ operators into their connected forms:
\begin{equation}
\label{eq:gpqrw3}
\begin{split}
G^R_{pq} (\omega) &=\\
&\left< \Phi \left| (1+\Lambda_2)
(a^{\dagger}_q + (a^{\dagger}_q T_2)_C)
X_{p}(\omega) \right| \Phi \right> +\\
&\left< \Phi \left| (1+\Lambda_2)
(a_p + (a_p T_2)_C)
Y_{q}(\omega) \right| \Phi \right>
\end{split}
\end{equation}
where the $X_p$ and $Y_q$ operators are
\begin{equation}
X_{p}(\omega) \left| \Phi \right> = (\omega+\bar{H}_N+i\delta)^{-1}
\bar{a}_p \left| \Phi \right>,
\end{equation}
\begin{equation}
Y_{q}(\omega) \left| \Phi \right> = (\omega-\bar{H}_N+i\delta)^{-1}
\bar{a}^{\dagger}_q \left| \Phi \right>,
\end{equation}
with components
\begin{equation}
X_{p}(\omega) = \sum_i x^i(\omega)_p a_i +
\frac{1}{2!}\sum_{ij,a} x^{ij}_a(\omega)_p a^{\dagger}_a a_j a_i = X_{1,p}(\omega)+X_{2,p}(\omega),
\end{equation}
\begin{equation}
Y_{q}(\omega) = \sum_a y_a(\omega)_q a^{\dagger}_a +
\frac{1}{2!}\sum_{i,ab} y^i_{ab}(\omega)_q a^{\dagger}_a a^{\dagger}_b a_i = Y_{1,q}(\omega)+Y_{2,q}(\omega).
\end{equation}
We proceed in the frequency domain because this simplifies the perturbation analysis. Each term in Eq.\ (\ref{eq:gpqrw3}) generates six possible terms, but, analyzing their corresponding diagrams we can see that there are only three non-zero terms when $p,q \in occ$:
\begin{equation}
\label{eq:gpqrw4}
G^R_{pq} (\omega) =
\left< \Phi \left| a^{\dagger}_q X_{1,p}(\omega) \right| \Phi \right> +
\left< \Phi \left| \Lambda_2 (a^{\dagger}_q T_2)_C
X_{1,p}(\omega) \right| \Phi \right> +
\left< \Phi \left| \Lambda_2 a_p Y_{2,q}(\omega) \right| \Phi \right>.
\end{equation}
After computing each term we get:
\begin{equation}
\label{eq:gpqrw5}
G^R_{pq} (\omega) =
x^q(\omega)_p -\frac{1}{2} \sum_{ijab} \lambda^{ab}_{ij} t^{qj}_{ab}
x^i(\omega)_p
-\frac{1}{2} \sum_{iab} \lambda^{ab}_{pi} y^{i}_{ab}(\omega)_q.
\end{equation}
This is the form of the Green's function which can be used to approximate $G^R_{pq}$ using perturbation theory. For this purpose we
write each of the coefficients in Eq.\ (\ref{eq:gpqrw5}) up to second order in a perturbation parameter $\lambda$, e.g. $t_{pq}^{rs} = t_{pq}^{(0)rs} + \lambda t_{pq}^{(1)rs} + \lambda^2 t_{pq}^{(2)rs}$, etc. By
keeping only terms up to second order we get:
\begin{equation}
\label{eq:g0pqrw1}
G^{(0)R}_{pq}(\omega) = x^{(0)q}(\omega)_p
\end{equation}
\begin{equation}
\label{eq:g1pqrw1}
G^{(1)R}_{pq}(\omega) = x^{(1)q}(\omega)_p
-\frac{1}{2} \sum_{iab} \lambda^{(1)ab}_{pi} y^{(0)i}_{ab}(\omega)_q,
\end{equation}
\begin{equation}
\label{eq:g2pqrw1}
G^{(2)R}_{pq}(\omega) =
x^{(2)q}(\omega)_p
-\frac{1}{2} \sum_{ijab} \lambda^{(1)ab}_{ij} t^{(1)qj}_{ab} x^{(0)i}(\omega)_p
-\frac{1}{2} \sum_{iab} (\lambda^{(1)ab}_{pi} y^{(1)i}_{ab}(\omega)_q +
\lambda^{(2)ab}_{pi} y^{(0)i}_{ab}(\omega)_q)
\end{equation}
It is easy to prove that $x^{(1)q}(\omega)_p=0$ and that
$x^{(0)q}(\omega)_p = x^{(0)p}(\omega)_p \delta_{pq}$. It can also be easily
proven that $Y^{(0)}_q(\omega) = 0$, resulting in $G^{(1)R}_{pq} (\omega) =0$.
After this, the retarded Green's function to second order is:
\begin{equation}
\label{eq:g2pqrw}
G^{R}_{pq} (\omega) =
x^{(0)p}(\omega)_p \delta_{pq} + x^{(2)q}(\omega)_p
-\frac{1}{2} \sum_{iab} \lambda^{(1)ab}_{pi}
(t^{(1)qi}_{ab} x^{(0)p}(\omega)_p + y^{(1)i}_{ab}(\omega)_q)
\end{equation}
Here we skip the derivation of each of the coefficients in term of the two-particle integrals and HF eigenvalues and simply list expressions:
\begin{equation}
\label{eq:x0w}
x^{(0)p}(\omega)_p = \frac{1}{(\omega-\epsilon_p)}
\end{equation}
\begin{equation}
\label{eq:x2w}
x^{(2)q}(\omega)_p = \frac{1}{2(\omega-\epsilon_q)}\left[
\sum_{ija} v^{qa}_{ij} x^{(1)ij}_{a}(\omega)_p + \right.
\left.\frac{1}{(\omega-\epsilon_p)}
\sum_{iab} v^{pi}_{ab} t^{(1)qi}_{ab} \right]
\end{equation}
\begin{equation}
\label{eq:x1w}
x^{(1)ij}_{a}(\omega)_p = \frac{v^{ij}_{pa}}{(\omega-\epsilon_p)
(\omega+\epsilon_a-\epsilon_i-\epsilon_j)}
\end{equation}
\begin{equation}
\label{eq:tl1}
t^{(1)ij}_{ab} = \lambda^{(1)ab}_{ij} = \frac{v^{ij}_{ab}}
{(\epsilon_i+\epsilon_j-\epsilon_a-\epsilon_b)}=
\frac{v^{ij}_{ab}}{\epsilon^{ij}_{ab}}
\end{equation}
\begin{equation}
\label{eq:x2w}
x^{(1)ij}_{a}(\omega)_p = \frac{v^{ij}_{pa}}{(\omega-\epsilon_p)
(\omega+\epsilon_a-\epsilon_i-\epsilon_j)}
\end{equation}
The final Dyson and cumulant form of $G^R_{pq}(t)$ becomes pparent after inserting these expressions into Eq. (\ref{eq:g2pqrw}) and proceeding as described in the main paper. 

\subsection{One-particle simplification of the EOM-CCS equations}

Given the computational demand of the full EOM-CCS method, it is of interest to see if approximations other than those explored in the main paper are possible. In particular, approximations arising from effective one-body Hamiltonians since these are common in the works of Hedin, Nozieres, Langreth, etc. The full EOM-CCS method has the following matrix element for the computation of the amplitude variation:

\begin{equation}
\label{eqn:matel2}
\begin{split}
\left< \phi_{i}^{a} \right| \bar{H}_N(t) \left| \phi \right> =&
f_{ai} + \sum_b f_{ab} t_i^b - \sum_j f_{ji} t_j^a - \sum_{jb} f_{jb} t_i^b t_j^a\\
& + \sum_{jb} v_{aj}^{ib} t_j^b
- \sum_{jkb} v_{ib}^{jk} t_j^a t_k^b + \sum_{jbc} v_{aj}^{bc} t_i^b t_j^c
- \sum_{jkbd} v_{jk}^{bd} t_i^b t_j^a t_k^d,
\end{split}
\end{equation}
where the $f$ terms come from the one-particle part of $H$ and the $v$ terms from the two-particle part. A simple, yet extreme approximation would be to only retain the first line in Eq. (\ref{eqn:matel2}). However, this approximation leaves out the important linear term in $v$. We now partition that term as follows:
\begin{equation}
\sum_{jb} v_{aj}^{ib} t_j^b = v_{ia}^{ia} t_i^a +
\sum_{j} v_{aj}^{ia} t_j^a + \sum_{b} v_{ai}^{ib} t_i^b +
\sum_{jb(\neq ia)} v_{aj}^{ib} t_j^b
\end{equation}
Discarding the last term and introducing the approximate sum into the matrix element expression above while only keeping the $f$ terms we obtain:
\begin{equation}
\label{eqn:matel3}
\left< \phi_{i}^{a} \right| \bar{H}_N(t) \left| \phi \right> =
f_{ai} + v_{ia}^{ia} t_i^a + \sum_b (f_{ab}-v_{aj}^{bi}) t_i^b
- \sum_j (f_{ji}-v_{aj}^{ia}) t_j^a
- \sum_{jb} f_{jb} t_i^b t_j^a
\end{equation}
Therefore, with this approximation the form of the one-particle $H$ is preserved, but with modified $f$ elements, which now include a correction for the particular $(i,a)$ valence-valence excitation. We are currently exploring the possibility of using this simplified propagation form with $f$ and $v$ parameters from effective Hamiltonians.

\section{Results}

\subsection{Effect of potential integral trimming}

\begin{figure}[t]
\includegraphics[scale=0.5,clip,
trim=1.1cm 1.2cm 3.0cm 1.8cm]{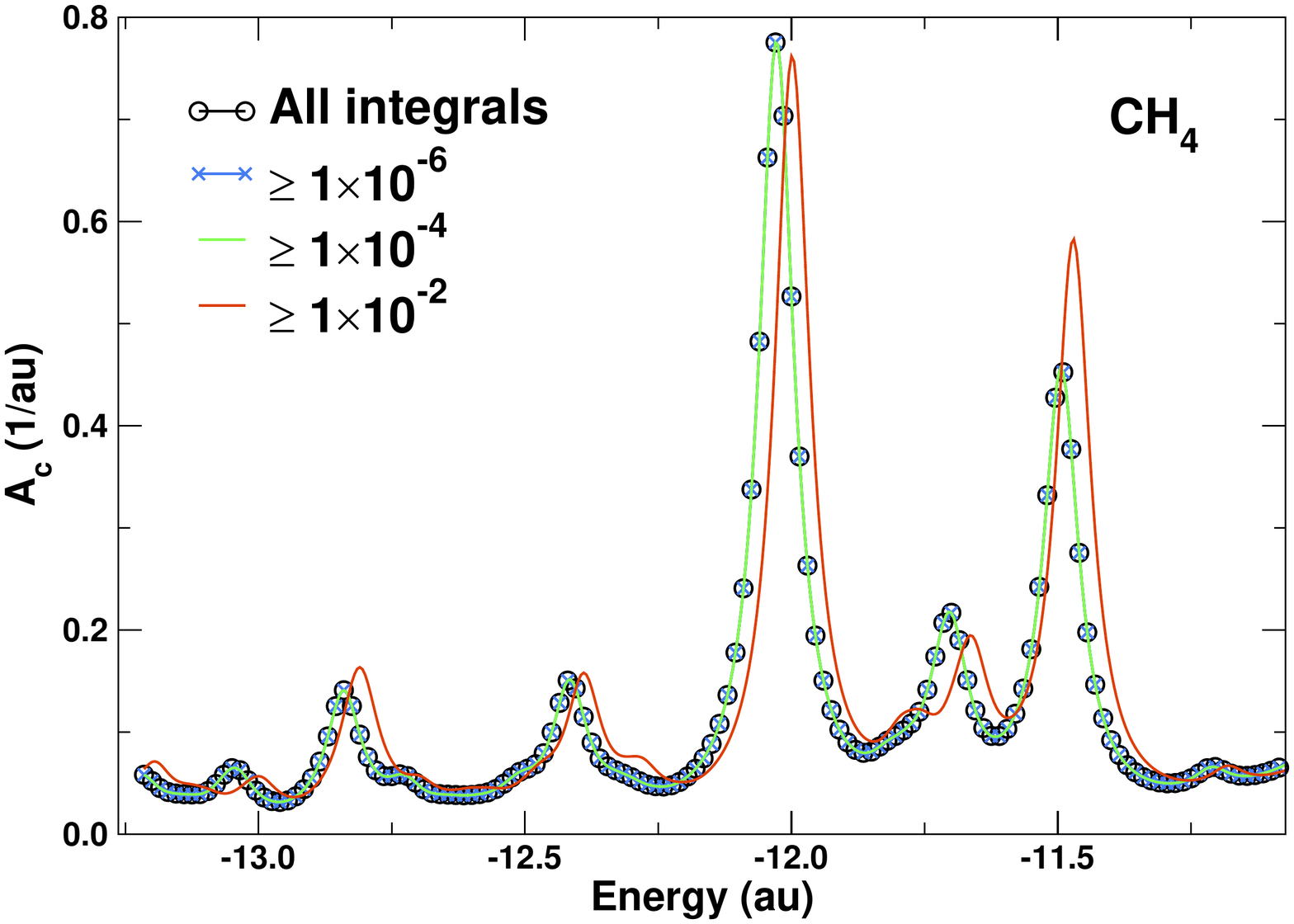}
\caption{\label{fig:cutoff}
Comparison of the satellite region of the core spectral function of CH$_4$ computed with the cc-pVDZ basis set and the full EOM-CCS method, as a function of potential integral cutoff.
}
\end{figure}
As discussed in the main manuscript, in order to reduce the storage requirements and computational demands of the EOM-CC method, we screen those $v_{pq}^{rs}$ integrals below a certain threshold after the SCF is fully converged. Figure \ref{fig:cutoff} shows a comparison between the spectral function of a typical system (CH$_4$) for different values of the cutoff parameter. We find that for the cutoff used in this paper (1$\times$10$^{-4}$ au) the EOM-CC results are indistinguishable from those obtained with all the integrals. With this cutoff, however, the performance of the method is increase by a factor of 10. This cutoff is also used in the DSE2 calculations, where it also has little effect in the accuracy, but produced only a modest improvement in performance. It should be noted that this integral trimming is not used in the GFCCSD and GFCC-i(2,3) calculations.

\subsection{Quasiparticle properties with the DZVP and cc-pVDZ basis sets}

Tables \ref{tbl:ebbas1} and \ref{tbl:ebbas2} summarize the core binding energies
for the different systems and methods calculated with the DZVP and cc-pVDZ basis
sets, respectively. While the KT, DSE2, and EOM-CC results show little
dependence on the basis set, for the GFCC methods the augmented Dunning basis
set seems not able to improve the results. To reduce the discrepancy of the GFCC
results, the employment of bare Dunning basis sets seems slightly better. For
the GFCCSD results, employing the aug-cc-pVDZ basis gives the MAE of 4.24 eV, in
comparison to the MAE of 3.57 eV brought by employing the bare cc-pVDZ basis.
Furthermore, the triple-$\zeta$ cc-pVTZ basis can systematically reduce the
discrepancies to even below 1 eV ($\sim$0.74 eV), which agrees with basis set
discussion in the previous EOM-CC and GFCC results for the core ionizations of
small molecules.\cite{mukherjee13_2625, sonia16_149901, peng18_4335}

\begin{table*}[t]
\caption{Comparison of the experimental core binding energies (in eV) to those obtained with the DZVP basis set, using the L and NL approximations to the cumulant and the 1-3 approximations of the EOM-CCS method, and their mean absolute errors (MAE).
}
\label{tbl:ebbas1}
\begin{ruledtabular}
\begin{tabular}{lddddddddddl}
\tbhr{System} & \tbhc{KT} &\tbhc{DSE2}& \tbhc{1$_{L}$} & \tbhc{2$_{L}$} & \tbhc{3$_{L}$} && \tbhc{1$_{NL}$} & \tbhc{2$_{NL}$} & \tbhc{3$_{NL}$} & \tbhc{Expt}& \tbhc{Ref} \\
\hline
CH$_4$&304.744&291.881& 286.990& 287.425& 286.994&& 290.412& 290.679& 290.415& 290.703&[\onlinecite{ch4spf}]\\
NH$_3$&422.523&405.466& 400.603& 400.815& 400.198&& 405.057& 405.177& 404.816& 405.520&[\onlinecite{nh3spf}]\\
H$_2$O&559.003&538.597& 534.795& 534.390& 533.705&& 539.498& 539.248& 538.843& 539.700&[\onlinecite{h2ospf}]\\
HF    &714.753&692.127& 689.876& 688.904& 688.313&& 694.174& 693.549& 693.178& 694.200&[\onlinecite{hfspf}] \\
Ne    &890.987&868.010& 867.661& 866.444& 866.109&& 870.935& 870.076& 869.842& 870.200&[\onlinecite{nespf}] \\
\hline
MAE   & 18.34 &1.32&   4.08 &   4.47 &   5.00 &&   0.34 &	0.32 &	 0.65 &        &		     \\
\end{tabular}
\end{ruledtabular}
\end{table*}
\begin{table*}[t]
\caption{Comparison of the experimental core binding energies (in eV) to those obtained with the cc-pVDZ basis set, using the L and NL approximations to the cumulant and the L$_c$, Q and F approximations of the EOM-CCS method, and their mean absolute errors (MAE). The KT results are obtained with Koopmans' Theorem.
}
\label{tbl:ebbas2}
\begin{ruledtabular}
\begin{tabular}{ldddddddddddl}
\tbhr{System} & \tbhc{KT} & \tbhc{DSE2}&\tbhc{\tiny{GFCCSD}} &\tbhc{\tiny{GFCC-i(2,3)}}  & \tbhc{1$_{NL}$} & \tbhc{2$_{L}$} & \tbhc{3$_{L}$} & \tbhc{1$_{L}$} & \tbhc{2$_{NL}$} & \tbhc{3$_{NL}$} & \tbhc{Expt}& \tbhc{Ref} \\
\hline
CH$_4$&305.17&292.56& 292.80 & 293.45 &286.98&287.84&287.44&290.54&291.08&290.83& 290.703&[\onlinecite{ch4spf}]\\
NH$_3$&422.78&406.26& 407.68 & 408.61 &400.67&401.35&400.81&405.13&405.55&405.23& 405.520&[\onlinecite{nh3spf}]\\
H$_2$O&559.25&539.30& 542.21 & 543.41 &534.53&534.74&534.15&539.32&539.46&539.10& 539.700&[\onlinecite{h2ospf}]\\
HF    &715.09&692.64& 696.78 & 698.19 &689.27&688.97&688.45&693.78&693.59&693.27& 694.200&[\onlinecite{hfspf}] \\
Ne    &891.59&868.17& 874.52 & 874.52 &866.50&865.87&865.57&870.16&869.73&869.52& 870.200&[\onlinecite{nespf}] \\
\hline
MAE   & 18.71 &1.32& 2.73& 3.57 &  4.48 &  4.31 &  4.78 &  0.28 &  0.35 &  0.53 &	 &		       \\
\end{tabular}
\end{ruledtabular}
\end{table*}

Tables \ref{tbl:zbas1} and \ref{tbl:zbas2} summarize the core quasiparticle strengths for the different systems and methods calculated with the DZVP and cc-pVDZ basis sets, respectively.
\begin{table*}[t]
\caption{Comparison of the quasiparticle strengths obtained with the DZVP basis set, using the L and NL approximations to the cumulant and the 1-3 approximations of the EOM-CCS method.
}
\label{tbl:zbas1}
\begin{ruledtabular}
\begin{tabular}{lddddddddddl}
\tbhr{System} & & \tbhc{DSE2} & \tbhc{1$_{L}$} & \tbhc{2$_{L}$} & \tbhc{3$_{L}$} && \tbhc{1$_{NL}$} & \tbhc{2$_{NL}$} & \tbhc{3$_{NL}$} &&\\
\hline
CH$_4$&&0.79&  0.60&  0.61&  0.59&&  0.70&  0.71&  0.69&&\\
NH$_3$&&0.77&  0.60&  0.61&  0.58&&  0.71&  0.71&  0.69&&\\
H$_2$O&&0.75&  0.63&  0.62&  0.59&&  0.73&  0.72&  0.70&&\\
HF    &&0.76&  0.68&  0.66&  0.64&&  0.76&  0.74&  0.72&&\\
Ne    &&0.78&  0.76&  0.73&  0.72&&  0.80&  0.78&  0.77&&\\
\end{tabular}
\end{ruledtabular}
\end{table*}
\begin{table*}[t]
\caption{Comparison of the quasiparticle strengths obtained with the cc-pVDZ basis set, using the L and NL approximations to the cumulant and the 1-3 approximations of the EOM-CCS method.
}
\label{tbl:zbas2}
\begin{ruledtabular}
\begin{tabular}{ldddddddddddd}
\tbhr{System} & \tbhc{DSE2}&\tbhc{GFCCSD} &\tbhc{GFCC-i(2,3)}  & \tbhc{1$_{L}$} & \tbhc{2$_{L}$} & \tbhc{3$_{L}$} && \tbhc{1$_{NL}$} & \tbhc{2$_{NL}$} & \tbhc{3$_{NL}$} &\\

\hline
CH$_4$&0.80& 0.77& 0.81& 0.60&  0.63&  0.61&&  0.70&  0.72&  0.71&&\\
NH$_3$&0.78& 0.76& 0.81& 0.61&  0.63&  0.61&&  0.71&  0.73&  0.71&&\\
H$_2$O&0.78& 0.71& 0.82& 0.64&  0.65&  0.63&&  0.74&  0.74&  0.72&&\\
HF    &0.79& 0.79& 0.84& 0.70&  0.69&  0.67&&  0.77&  0.76&  0.75&&\\
Ne    &0.81& 1.00& 1.00& 0.77&  0.76&  0.75&&  0.82&  0.81&  0.80&&\\
\end{tabular}
\end{ruledtabular}
\end{table*}
\begin{table}[t]
\caption{Scissors corrections used in Fig. \ref{fig:ac_sci}.}
\label{tbl:scicorr}
\begin{ruledtabular}
\begin{tabular}{lddddddl}
\tbhr{System} & &  & \tbhc{1$_{NL}$} & \tbhc{2$_{NL}$} & & &\\
\hline
CH$_4$&&&     1.9 &1.1 &  &  &\\
NH$_3$&&&     6.2 &1.6 &  &  &\\
H$_2$O&&&     8.8 &1.8 &  &  &\\
HF    &&&    10.4 &2.1 &  &  &\\
Ne    &&&    10.2 &1.5 &  &  &\\
\end{tabular}
\end{ruledtabular}
\end{table}

\subsection{Full comparison of the spectral function as a function of basis set, level of CCS approximation and cumulant form}

\begin{figure}[t]
\includegraphics[scale=0.40,clip]{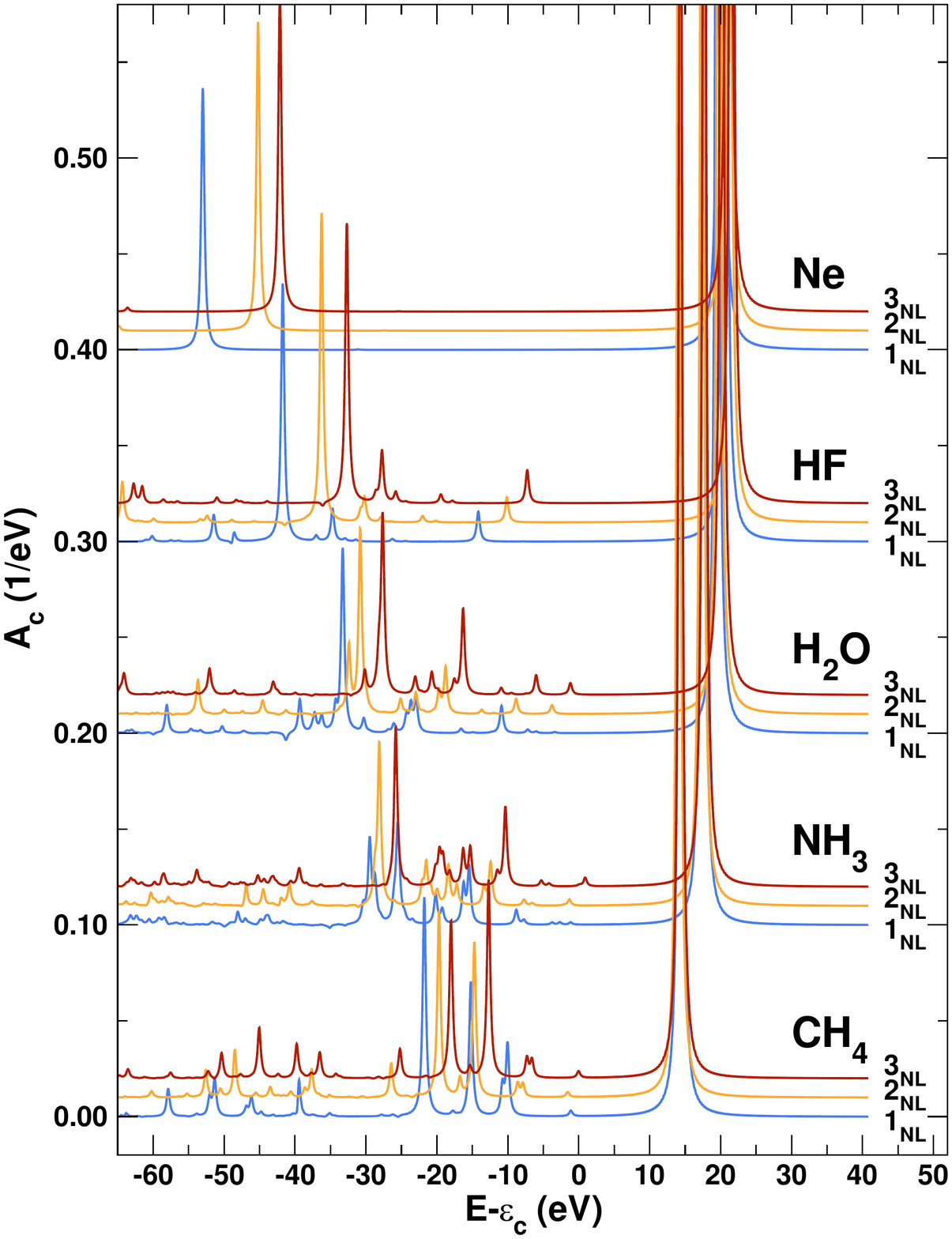}
\caption{\label{fig:acbas1}
Core spectral function $A_c$ of the $10e$ systems computed with the DZVP basis set, as a function of EOM-CCS approximation 1$_{NL}$ (blue), 2$_{NL}$ (orange), and 3$_{NL}$ (red).
}
\end{figure}

\begin{figure}[t]
\includegraphics[scale=0.40,clip]{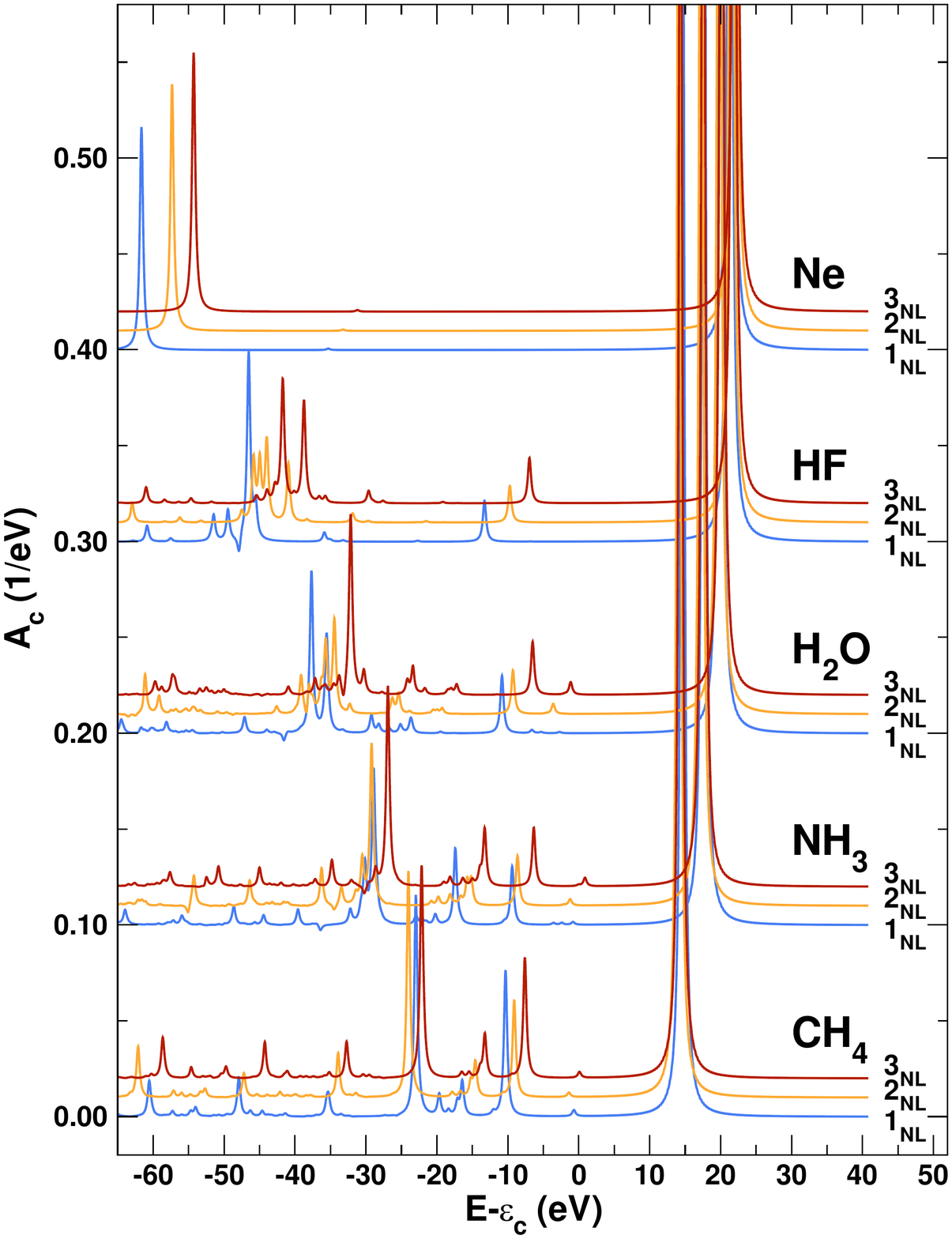}
\caption{\label{fig:acbas2}
Core spectral function $A_c$ of the $10e$ systems computed with the cc-pVDZ basis set, as a function of EOM-CCS approximation 1$_{NL}$ (blue), 2$_{NL}$ (orange), and 3$_{NL}$ (red).
}
\end{figure}

Figures \ref{fig:acbas1} and \ref{fig:acbas2} shows a comparison of the core spectral function of the $10e$ systems computed with the DZVP and cc-pVDZ basis sets, respectively, and the 3$_{NL}$ approach, as a function of EOM-CCS approximation.

\subsection{Effect of the CCS approach on the gap between the quasiparticle and the satellites}

\begin{figure}[t]
\includegraphics[scale=0.40,clip]{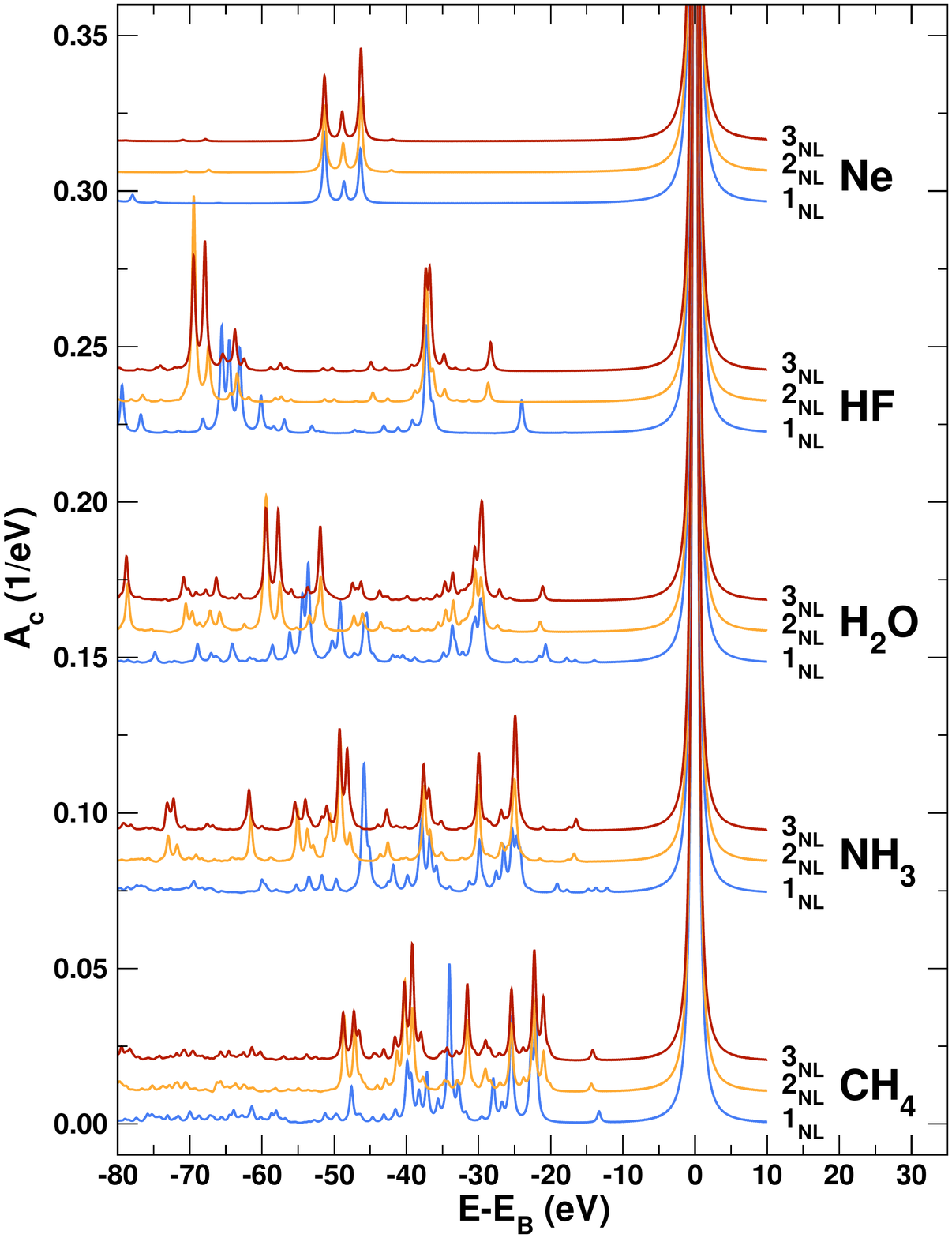}
\caption{\label{fig:ac_sci}
Comparison of the spectral functions with a scissors correction (see Table \ref{tbl:scicorr}) applied to the 1$_{NL}$ (blue) and 2$_{NL}$ (orange) approximations in such a way that the satellite regions become aligned with those of 3$_{NL}$ (red).
}
\end{figure}

Figure \ref{fig:ac_sci} shows a comparison of the spectral functions with a scissors correction applied to the 1$_{NL}$ and 2$_{NL}$ approaches in such a way that the satellite regions become aligned with those in the 3$_{NL}$ approach. After the correction is applied, the 2$_{NL}$ approximation is shown to give results that are almost identical to the full approach. Despite showing some noticeable differences, 1$_{NL}$ approximation shows reasonable agreement in the overall distribution of the satellite intensity. The scissors corrections for each system and method are shown in Table \ref{tbl:scicorr}. Interestingly, while the corrections required for the 1$_{NL}$ approximation are clearly system-dependent, in the case of the 2$_{NL}$ approximation the corrections are almost constant, suggesting that an overall scissors correction can be used to simulate the results of the more expensive full approach.

\bibliography{EOM-CC-Cumulant-Approach}
\bibliographystyle{apsrev}